\documentclass[usenatbib]{mn2e}
\usepackage{bm}
\usepackage{morefloats}
\usepackage{hyperref}
\usepackage{tabularx}
\usepackage{graphicx}
\usepackage{multirow}
\usepackage{pdflscape}
\usepackage{amsmath,amssymb}

\newcommand {\gaia}{\textit{Gaia }}

\title[Evidence for a bottom-light IMF in star clusters]
{Evidence for a bottom-light initial mass function in massive star clusters}
\author[Baumgardt et al.]{H. Baumgardt$^{1}$\thanks{E-mail: h.baumgardt@uq.edu.au}, V. H\'enault-Brunet$^{2}$, N. Dickson$^{2}$, A. Sollima$^{3}$\thanks{Deceased}\\
$^{1}$ School of Mathematics and Physics, The University of Queensland, St. Lucia, QLD 4072, Australia \\
$^{2}$ Department of Astronomy and Physics, Saint Mary's University, 923 Robie Street, Halifax, NS B3H 3C3, Canada\\
$^{3}$ INAF Osservatorio Astronomico di Bologna, via Gobetti 93/3, Bologna, 40129, Italy\\
}

\begin{document}

\date{Accepted 2021 xx xx. Received 2021 xx xx; in original form 2021 xx xx}

\pagerange{\pageref{firstpage}--\pageref{lastpage}} \pubyear{201x}

\maketitle

\label{firstpage}

\begin{abstract}
We have determined stellar mass functions of 120 Milky Way globular clusters and massive LMC/SMC star clusters based on 
a comparison of archival Hubble Space Telescope photometry with a large grid of direct $N$-body simulations.
We find a strong correlation of the global mass function slopes of star clusters with both their internal relaxation times as well as their 
lifetimes. Once dynamical effects are being accounted for, the mass functions of most star clusters are compatible with an initial mass function
described by a broken power-law distribution $N(m) \sim m^{\alpha}$ with break masses at 0.4 M$_\odot$ and 1.0 M$_\odot$ and mass function
slopes of $\alpha_{Low}=-0.3$ for stars with masses $m<0.4$ M$_\odot$, $\alpha_{High}=-2.30$ for stars with $m>1.0$ M$_\odot$ and
 $\alpha_{Med}=-1.65$ for intermediate-mass stars. Alternatively, a log-normal mass function with a characteristic mass 
$\log M_C = -0.36$ and width $\sigma_C=0.28$ for low-mass stars and a power-law mass function for stars with $m>1$ M$_\odot$ also
fits our data. We do not find a significant environmental dependency of the initial mass function with either cluster mass, density, global
velocity dispersion or metallicity. Our results lead to a larger fraction of high-mass stars in globular clusters
compared to canonical Kroupa/Chabrier mass functions, increasing the efficiency of self-enrichment in clusters and helping to alleviate the mass 
budget problem of multiple stellar populations in globular clusters. By comparing our results with direct $N$-body simulations we finally find 
that only simulations in which most black holes are ejected by natal birth kicks correctly reproduce the observed correlations.
\end{abstract}

\begin{keywords}
globular clusters: general -- stars: luminosity function, mass function
\end{keywords}

\section{Introduction} \label{sec:intro}

The initial mass function (IMF) of stars is important in understanding a large number of astronomical phenomena such as the
formation of the first stars \citep{brommetal2009}, galaxy formation and evolution \citep[e.g][]{caluramenci2009,abeetal2021}, 
and the determination of the absolute star formation rate \citep{aoyamaetal2021}. It also plays a dominant role in any star formation theory as
the end result of molecular cloud contraction and fragmentation \citep[e.g.][]{krumholz2014}.

The stellar IMF was first measured by \citet{salpeter1955}, who found that the mass function of massive, $m>1$ M$_\odot$ stars 
in the solar neighborhood is best described by a power-law mass function $N(m) \sim m^{\alpha}$ with slope $\alpha=-2.35$.
Evidence has been accumulating that the mass function for lower-mass stars in the Galactic disc increases less strongly with decreasing
mass \citep{kroupa2001,chabrier2003}, but its exact form and whether it varies between individual star forming clouds is still under debate \citep[see review by][]{bastianetal2010}.

There has also been increasing evidence that the stellar mass function is varying with galaxy environment or cosmic time. The best 
possible case for IMF variations are probably the centers of early type galaxies, in which spectroscopic measurements \citep[e.g.][]{vandokkumconroy2010} as well as measurements
of the stellar kinematics \citep{cappellarietal2012} indicate that the low-mass star IMF must be bottom-heavy. It has also been suggested that, due to inefficient cooling,
the mass function of the first stars in the universe must have been top-heavy \citep{abeletal2002,brommetal2002}. Top-heavy IMFs for high mass stars have also been found
in massive Galactic molecular cloud complexes like W43 \citep{pouteauetal2022,nonyetal2023}, which could indicate that the IMF slope depends on star formation rate and that
starburst events create top-heavy IMFs.  Finally, theoretical arguments and radiation-hydrodynamical
simulations \citep[e.g.][]{krumholzetal2010} indicate that radiation feedback from forming stars or cooling from dust-grains \citet{chonetal2021} could influence
proto-stellar fragmentation and thereby the form of the IMF.

Star clusters are one of the best environments to determine the initial stellar mass functions since they offer large, statistical significant numbers of
stars of similar distance, age and chemical composition. The strong crowding of stars in the cluster centres makes the detection of their lowest mass
stars a challenge even when using space-based observatories like the Hubble Space Telescope (HST). In addition, the large angular extent of nearby star clusters
means that it is usually not possible to obtain photometry of the whole cluster with a single HST pointing. It is therefore necessary to use models like 
multi-mass King-Michie models \citep{gunngriffin1979,sollimaetal2017} that can correct for internal mass segregation in star clusters to
correct locally measured mass functions to the global mass function. In addition, the dynamical evolution of star clusters needs to be taken into account 
when trying to deduce the initial mass function of a cluster from the present-day one since
star clusters lose preferentially their lowest mass stars over time \citep{vesperiniheggie1997,baumgardtmakino2003}.

In this paper we determine the stellar mass functions of 120 Galactic globular clusters and Large and Small Magellanic cloud clusters from archival
HST photometry, obtaining the largest database of mass function measurements for these systems. We then use this data to determine their initial
mass functions and compare them to stellar mass functions measured for nearby galaxies.
Our paper is organised as follows. In Sect.~\ref{sec:phot} we describe the selection of the clusters and the analysis of the HST data and in Sect.~\ref{sec:mfunc}
we describe the determination of the stellar mass functions. In Sect.~\ref{sec:results} we describe the results and we draw our conclusions in
Sect.~\ref{sec:conclusions}.

\nocite{kamannetal2018} 
\nocite{vasilievbaumgardt2021}

\section{Photometry} \label{sec:phot}

We took the input list of Milky Way globular clusters from the most recent version of the globular cluster database of
\citet{baumgardtetal2019a}, which lists 165 Galactic globular clusters. From this list
we removed all clusters which either had no
existing deep HST photometry reaching several magnitudes below the main sequence turn-over, were in fields of high stellar background density, or had large extinction values $E(B-V)>0.80$.
We also removed clusters for which the available HST photometry was not deep enough to allow us to determine the stellar
mass function down to masses of at least $\sim$0.50 M$_\odot$. In total we found 91 Galactic globular clusters which fulfilled all of the above constraints
and we list these clusters in Table~\ref{table:main}. In order to extend the measured mass functions to stars with masses above 0.8~M$_\odot$, which
have already turned into compact remnants in $\sim$12 Gyr old globular clusters,
we also analysed stellar mass functions for 29 massive star clusters of the Large (LMC) and Small Magellanic Clouds (SMC) that have deep HST photometry
and we list the adopted parameters and derived mass function slopes of these clusters in Table~\ref{table:lmcsmc}.

For each star cluster we selected from the {\tt STSci} data archive suitable HST photometry, making sure that we could get 
an as large as possible radial range for which we can measure the stellar mass function. Due to their proximity, this generally required
us to analyse more than one HST field for Galactic globular clusters, while the       
more distant star clusters in the LMC and SMC usually fitted into a single HST field. 
Figs.~A1 to A20 depict for each star cluster the location of the analysed HST fields.

After downloading the HST data, we prepared the photometric images using the {\tt splitgroups} and camera-specific masking 
tasks as described in the 
{\tt DOLPHOT} handbook and then performed stellar photometry on the data using {\tt DOLPHOT} 
\citep{dolphin2000, dolphin2016}. For ACS and WFC3 data, we performed the photometry on the CTE corrected flc images, 
while for the WFPC2 observations we used the c0m images to perform the photometry. We used the point-spread functions provided for 
each camera and filter combination by {\tt DOLPHOT} for the photometric reductions. Photometry was done by using the drizzled drc and
drz images provided by the {\tt STSci} data archive as master frames, which also correct for geometric camera distortions. After obtaining the 
photometry, we removed detected objects that either had sharpness values $|s|>0.1$ or roundness parameters $r$ larger than $r>0.25$ from the list of sources. 
The final magnitude and their associated errors for each star were calculated as the average and the r.m.s. of the individual magnitudes.

After performing the photometry, we cross-matched the HST coordinates of bright stars with the positions 
of stars in the \gaia DR3 catalogue \citep{gaiadr3main} using the \gaia proper motions to move the 
\gaia positions from the 2016.0 \gaia DR3 epoch to the observation epoch of the HST data. We then applied position dependent shifts 
to the HST coordinates to bring them into agreement with the \gaia DR3 ones. These shifts were for each star calculated by determining
the median shift of the nearest 30 stars against their \gaia DR3 counterparts.
For clusters with a high
stellar background density, and if more than one data set in a given field was available, we performed astrometry for multiple epochs 
this way and then determined individual proper motions for the stars. We then fitted a Gaussian mixture model to the resulting proper
motion vector point diagram, modeling both the cluster and the background stars as two-dimensional Gaussians. From the fit we then calculated
membership probabilities and selected as cluster stars all stars that
had a probability larger than 10\% to be cluster members. We chose this relatively low limit since we can remove most of the remaining non-members
by the isochrone fits described further below. For clusters with significant stellar background density but only one available epoch, proper
motion cleaning is not possible. In order to remove the contribution of non-members in these clusters, we shifted the
best-fitting isochrone in colour and determined stellar number counts to the left and right of the cluster main sequence and in regions that are occupied only by background
stars. We then averaged the resulting stellar numbers per magnitude interval and subtracted them from the stellar number counts for the cluster main sequence.

\begin{figure*}
\begin{center}
\includegraphics[width=0.91\textwidth]{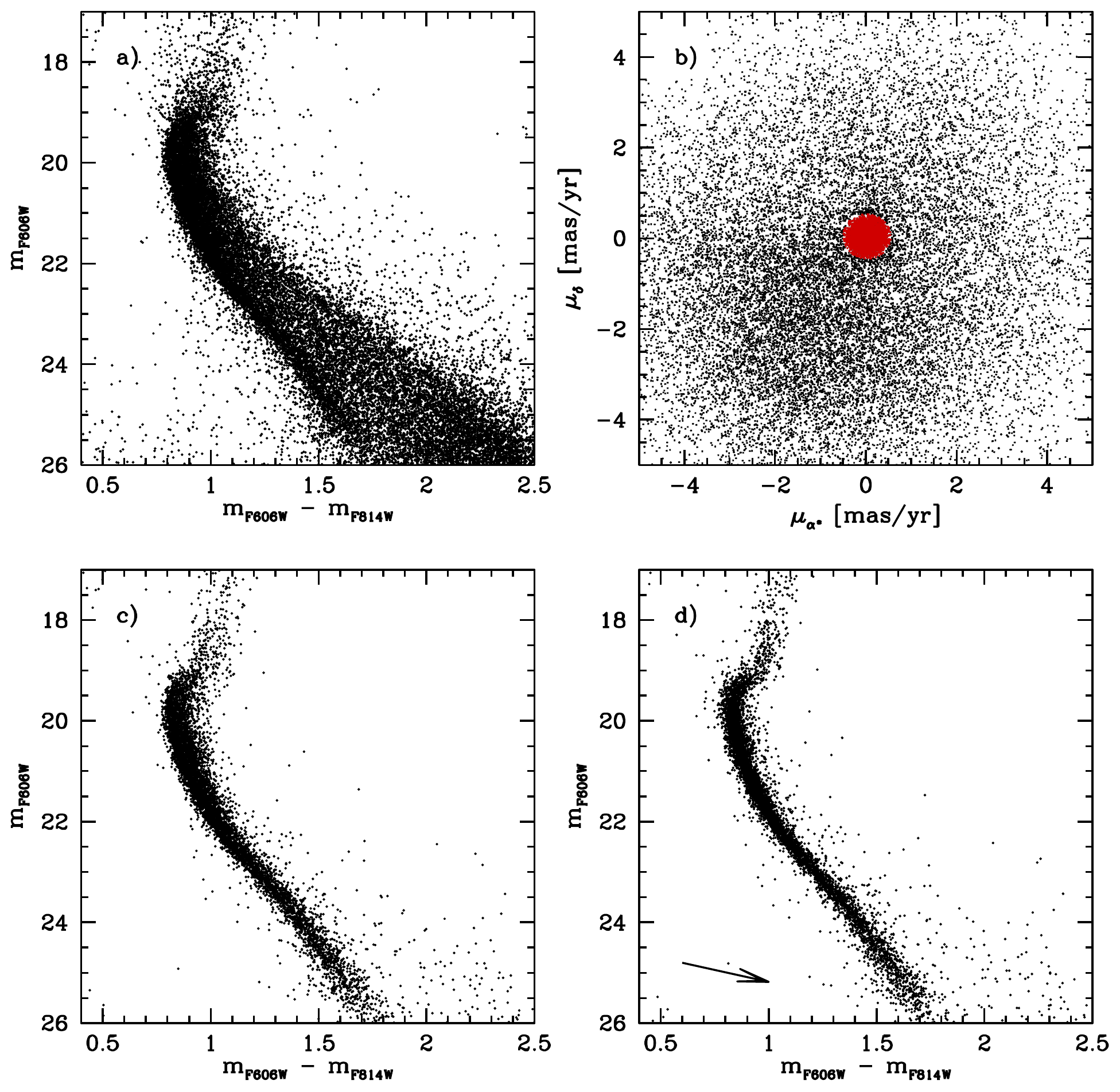}
\end{center}
\vspace*{-0.5cm}
\caption{Example showing the effects of proper motion cleaning and differential reddening correction on the observed CMD of NGC~6558. Panel a) shows an $m_{F606W}-m_{F814W}$
 vs. $m_{F606W}$ colour-magnitude diagram of all stars that pass the sharpness and roundness criteria. Panel b) shows the proper-motion vector point diagram of these stars
  with the cluster centered at the origin. Stars marked in red are stars that pass the proper motion membership test. Panel c) shows the CMD after stars
   with membership probabilities less than 10\% have been removed. Panel d) shows the CMD of the member stars after differential reddening correction, which leads to a 
    further narrowing of the CMD. The arrow in the bottom right panel shows the direction of the reddening vector.}
\label{fig:pmcleaning}
\end{figure*}

For clusters with significant and position dependent reddening we also de-reddened the final colour magnitude diagrams (CMDs) before
fitting the CMDs with stellar isochrones. Cluster de-reddening was done by first fitting an isochrone to the CMD of the central cluster parts. For each star
we then calculated its displacement from this isochrone along the reddening vector. The coefficients of the reddening vector were calculated based on the
analytic formulae of \citet{cardellietal1989} assuming $R(V)=3.1$. We also selected a magnitude interval where the CMD was
dominated by cluster stars. We then corrected the CMD position of each star by calculating the mean displacement of the stars nearest to it.
The number of stars used and the magnitude limits were varied for each cluster depending on the total number of cluster stars and the strength 
of the background contamination. Fig.~\ref{fig:pmcleaning} shows as an example the effects of proper motion 
cleaning and de-reddening for the globular cluster NGC~6558.

For each analysed HST field we also estimated the photometric completeness using artificial star tests. To this end, we distributed artificial stars with uniform 
spatial density across each HST field. Stars were equally spread in magnitude along the location of each cluster's main sequence from the 
turnover down to the faintest detectable magnitudes. We created $100,000$ artificial stars in HST fields that covered cluster centres and
$25,000$ stars for HST fields that covered areas outside the centre. We used a larger number of stars for central fields since these are more
affected by crowding and the completeness fraction will vary more quickly with radius since the stellar density varies strongly with radius in 
the centre. We used the {\tt DOLPHOT} 
{\tt fakestars} task to recover the magnitudes of the artificial stars and applied the same quality cuts to the artificial stars that we used to select 
real stars in the observed data sets. We then estimated the completeness fraction for each observed star from the ratio of successfully recovered stars to
all inserted stars using the nearest 20 artificial stars that are within 0.2 mag of the magnitude of each observed star. In order to limit the
influence of photometric incompleteness, we analysed mass functions only down to magnitudes where the average completeness is above 75\% in each field.
The only exception were WFPC2 observations where {\tt DOLPHOT} seemed to have problems in properly aligning the individual data frames to the drizzled {\tt drz}
master frames and in which the photometric completeness was typically only around 75\% even for bright, non-saturated stars. Since the alignment problems do 
not seem to depend on the stellar magnitudes and the completeness tests seem to be able to correct for their effect, we adopted
a smaller completeness limit of 50\% for WFPC2 data.

\section{Mass function determination} \label{sec:mfunc}

\subsection{Isochrone fits}

We determined stellar mass functions by performing isochrone fits to the observed CMDs. In order to determine the best-fitting isochrone, we varied the assumed 
cluster distances and reddenings but took the cluster ages and metallicities for which the isochrones were generated from literature data. We took the ages mainly
from the compilations of \citet{deangelietal2005}, \citet{marinfranchetal2009}, \citet{dotteretal2010}, \citet{vandenbergetal2013} and \citet{valcinetal2020}. 
In order to account for systematic differences between the ages derived in these papers, we first averaged all ages and then calculated
shifts for each individual paper against the mean ages determined this way. For the comparison we allowed the shifts to depend linearly on metallicity. 
This assumption is compatible with the observed age differences between different studies, see e.g. Fig.~9 in \citep{marinfranchetal2009}.
After correcting systematic differences in this way, we then calculated a final average age of each cluster. We took the cluster metallicities 
from the compilation of \citet{carrettaetal2009b}. 
We then created DSEP isochrones \citep{dotteretal2008} with these metallicities and ages and fitted the isochrones to the cluster CMDs. We used DSEP
isochrones with an $\alpha$ element abundance enhancement of $\alpha=0.2$ for clusters with [Fe/H]$<-1.0$ and solar composition for clusters 
with metallicities larger than [Fe/H]$=-1.0$. To test
the influence of the choice of isochrones on our results, we also fitted PARSEC isochrones \citep{bressanetal2012,chenetal2014,chenetal2015} to the clusters, but found
that the derived global mass function slopes changed by less than $\Delta \alpha = 0.20$ dex.

In order to fit the isochrones, we took the initial values of the cluster reddening from \citet{harris1996} and the cluster distances from \citet{baumgardtvasiliev2021}. 
We then varied reddening
and distance until we achieved the best fit to each observed CMD. Table~\ref{table:main} gives the best-fitting reddenings and distances for each cluster,
averaged over all individual CMDs that we fitted for each cluster. The resulting reddening and distance values are 
usually within $\pm 0.05$ mag of the reddening given by \citet{harris1996} and within $\pm 0.10$ in distance modulus to the distances from \citet{baumgardtvasiliev2021}. 
Once the best-fitting isochrones had been determined, we selected all stars with photometry compatible with each isochrone in radial annuli of 20'' width. 
We also calculated a best-fitting power-law mass function slope $N(m) \sim m^{\alpha}$ for each radial annulus 
using the maximum likelihood method described in \citet{clausetetal2009} and \citet{khalajbaumgardt2013}, which uses unbinned data. We note
that our final results are rather insensitive to the assumed ages and metallicities, the measured mass function
slope changes for example by only about 0.10 dex for a change in cluster age of 1 Gyr or a change in metallicity by $\Delta$[Fe/H]=0.20.

\section{Results} \label{sec:results}

\subsection{Initial mass function of clusters}
\label{sec:imf}

In this section we derive the mass function of star clusters that have relaxation times $T_{RH}$ of the order of their ages and total lifetimes 
at least three times larger than their ages.
These clusters should have largely preserved their mass functions since they lost only a small fraction of their mass. In addition, due to
their large relaxation times such clusters are also not strongly mass segregated and therefore even the small amount of mass loss they have experienced will 
lead to a loss of stars independent of their masses (at least as long as the clusters started without primordial mass segregation). We therefore assume that 
the present-day mass functions of these clusters resemble their initial mass functions. These assumptions are confirmed by the results of the $N$-body simulations 
done in sec.~4.2 of this paper as well as by the $N$-body simulations done by \citet{webbleigh2015}. We derive the mass function for stars with masses $m<0.8$ M$_\odot$
from Milky Way globular clusters and the mass function of higher mass stars from the LMC and SMC star clusters.

\subsubsection{The low-mass star IMF}
\label{sec:lowmassmf}

In this section we restrict ourselves to globular clusters that have mass function determinations both inside and outside their half-mass radii and for which 
the mass functions can be determined down 
to at least $m=0.25$ M$_\odot$. This way we can measure the mass function directly from the observations and are independent of model fits.
These requirements restrict us to six globular clusters: IC~4499, NGC~5024, NGC~5053, NGC~5139, NGC~6101 and Ter~8.
\begin{figure*}
\begin{center}
\includegraphics[width=0.97\textwidth]{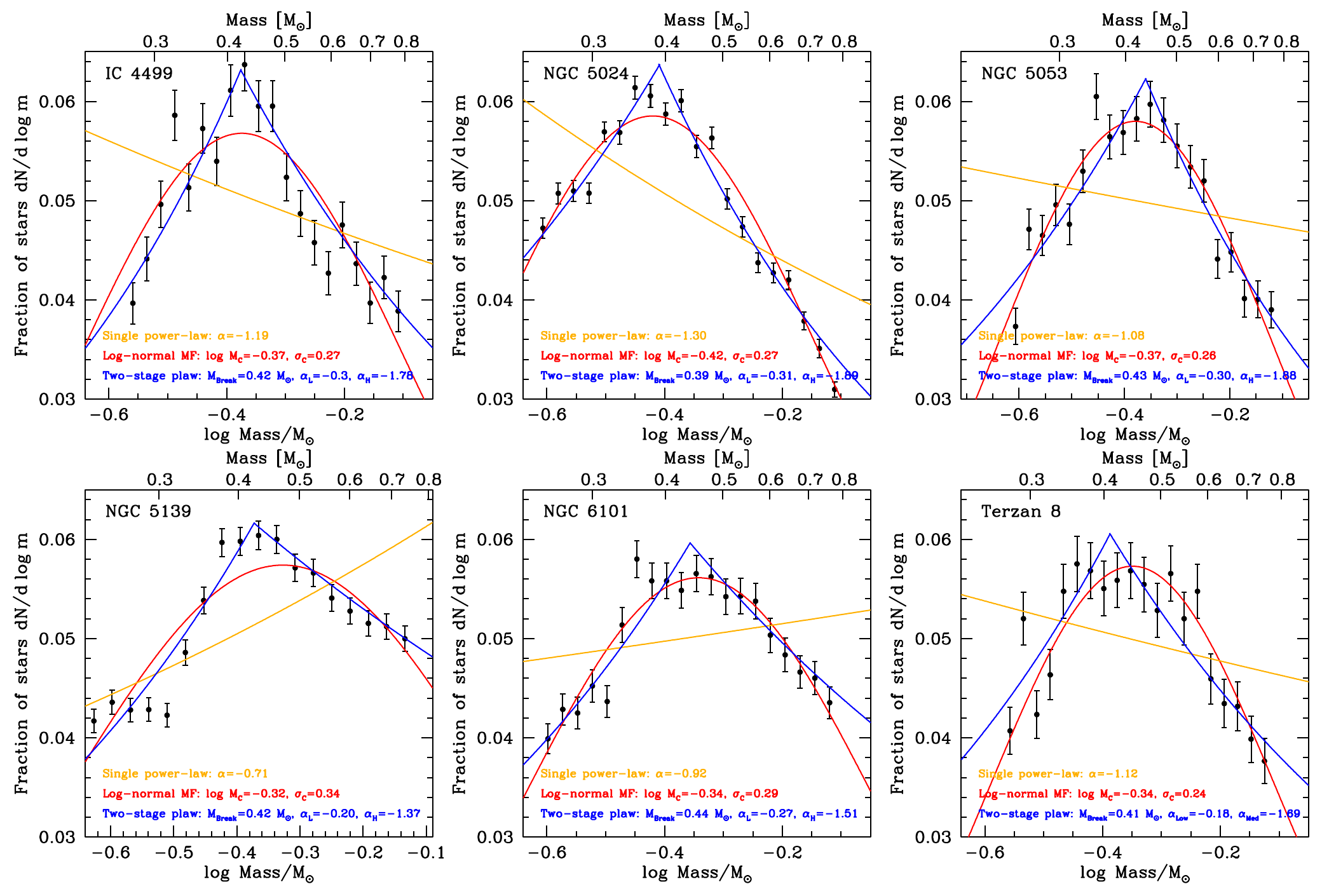}
\end{center}
\vspace*{-0.2cm}
\caption{Global mass functions for (clockwise from top left) IC~4499, NGC 5024, NGC 5053, Ter~8, NGC 6101 and NGC 5139, the six globular clusters in our sample that
have both large relaxation and lifetimes and photometry deep enough to reach down to at least 0.25~M$_\odot$. Filled black circles with error bars show the fraction of
observed stars as a function of mass in the different clusters and their 1$\sigma$ errors. The orange, blue and red lines show fits
of the best-fitting single power law, two-stage power law and log-normal mass functions to the data. The parameter of the fits are given in each panel.
There is generally quite good agreement between the parameters that we find for each cluster, implying that all clusters have started with very similar mass functions.
A two-stage power-law and a log-normal mass function provide about equally good fits to the data, while a single power-law mass function overpredicts the
number of high and low-mass stars and underpredicts the number of stars with masses around 0.4 M$_\odot$. The deviations are however only of order 10\%.}
\label{fig:mfall}
\end{figure*}

For each cluster we use the method described in \citet{baumgardtetal2022} to determine the fraction $f_r$ of the cluster that is covered by our photometry at the projected distance 
of each star from the centre. We then assign a weighting factor $w=1/(f_r f_p)$ to each star where $f_p$ is the fraction of stars of similar magnitudes and location that are 
recovered in our completeness tests. For NGC~5024 and NGC~5139 we are not able
to apply this procedure to the innermost parts since our photometry becomes incomplete at masses $m \sim 0.5$ M$_\odot$ due to crowding. The derived mass functions
for these clusters could therefore be slightly skewed towards lower masses as we tend to miss preferentially higher mass stars due to mass segregation. However
the effect is likely only small since
we are analysing clusters with large relaxation times which are not strongly mass segregated. We similarly lose a few percent of the outermost stars in all
clusters since our photometry does not reach the tidal radius, however we again expect that this is not going to have a strong effect on our results.
\begin{figure*}
\begin{center}
\includegraphics[width=0.95\textwidth]{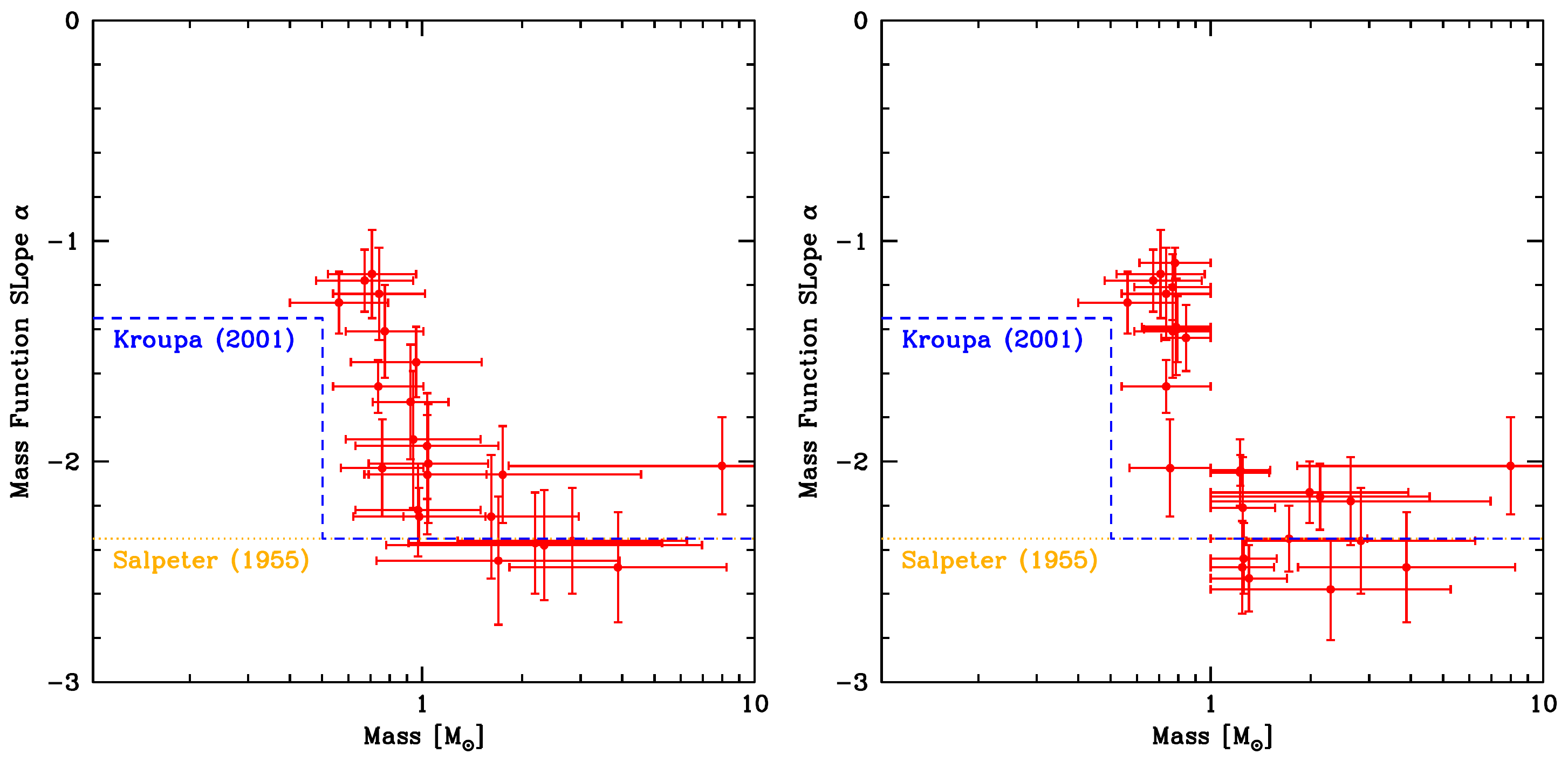}
\end{center}
\vspace*{-0.2cm}
\caption{Mass function slopes for massive LMC and SMC star clusters. Horizontal lines depict the mass range for which the mass functions have been determined, vertical lines
 depict the uncertainty in the derived mass function slopes. In the left panel we show the slopes derived over the full mass range that could be fitted in each cluster,
 while in the right panel we have split the sample into two mass ranges for stars less and more massive than 1 M$_\odot$. There seems to be a clear break in mass
  function slope at around 1 M$_\odot$.}
\label{fig:lmcsmcmf}
\end{figure*}

After deriving the individual stellar masses, we then fit a single-power mass function, a two-stage power-law mass function
with a break-mass $m_{Break}$ and a mass function slope $\alpha_{Med}$ for stars with $m>m_{Break}$ and a low-mass slope $\alpha_{Low}$ for stars with $m<m_{Break}$,
and a log-normal mass function $\phi(m) \sim e^{-(log M_C-log m)^2/(2 \sigma_C^2)}$ with a characteristic mass $M_C$ and a width $\sigma_C$ to the data.
We derive the best-fitting parameters for each of these models using a maximum-likelihood approach,
following the procedure outlined by \citet{clausetetal2009}.  Fig.~\ref{fig:mfall} depicts the mass distribution of stars in each of the six clusters as
well as the results of our fits. It can be seen that single power-law mass functions are not accurate fits to the data since they predict too many
high and low-mass stars and too few intermediate stars with masses around 0.4 M$_\odot$. However, the relative deviations from the actual data are for most masses
only of order 10\%, so a single power-law mass function is still a useful approximation. Fig.~\ref{fig:mfall} also shows that two-stage power-law mass functions
and log-normal mass functions provide significantly better fits to the data. For a two-stage
power-law mass function, we obtain break masses $m_{Break}$ between 0.39 M$_\odot$ to 0.44 M$_\odot$, mass function slopes between $\alpha_M \approx -1.40$ to
$\alpha_M \approx -1.90$ for stars more massive than the break mass and significantly flatter mass function slopes for the low-mass stars. The mass function does therefore 
flatten towards lower masses, a behavior qualitatively similar to that seen for Galactic
disc stars \citep{kroupa2001}. The individual slopes are however flatter at all masses, leading again to a mass function with a smaller fraction of low-mass stars
compared to what is found in the Galactic field.
Taking the average over all six clusters we obtain for the break mass, and the high and low-mass slopes:  $m_{Break} = 0.42 \pm 0.02$, $\alpha_{Med}=-1.68 \pm 0.20$ and
$\alpha_{Low}=-0.27 \pm 0.08$. Here the error bars reflect the standard deviation of the individual clusters around the mean. We decided to use the standard deviation 
as an estimate of the uncertainty since the formal error bars are usually small (typically of order 0.02) and do not reflect systematic errors which, under the assumption
of a common initial mass function, should be better represented by the standard deviation.

Fig.~\ref{fig:mfall} also shows that log-normal mass functions provide acceptable fits to the data. The values that we derive for the parameters are again 
similar between the different clusters, making it possible that all clusters have started with the same mass function. Taking an average over all clusters, we find
$\log M_C = -0.36 \pm 0.03$ and $\sigma_C = 0.28 \pm 0.04$. The \citet{chabrier2003} mass function, which fits the distribution of stars in the Galactic disc, has a characteristic
mass of $\log M_C = -1.00$. Hence our results again argue for a smaller fraction of low-mass stars in globular clusters compared to the Galactic disc.

\subsubsection{The high-mass star IMF}
\label{sec:highmassmf}

Since globular clusters only allow to determine the sub-solar stellar mass function, we next extend our analysis to the stellar mass functions of a number of 
massive star clusters in the LMC/SMC that have available deep HST photometry. The chosen clusters span a range of ages between 3 Myr and 12 Gyr and have masses from about
$10^4$~M$_\odot$ to $7 \cdot 10^5$ M$_\odot$, similar to the masses of the globular clusters studied previously.  We analyse each LMC/SMC star cluster in the same way as 
the Milky Way globular clusters by deriving their photometry from HST observations. 
Due to the large distances of the LMC and SMC, a single HST field is usually sufficient to cover most stars in a star cluster. 
From a comparison of the available observational data of each cluster to a grid of $N$-body simulations (to be described below), we then derive the physical parameters of the clusters
including their mass functions. The basic data of the clusters (distances, ages, extinctions) are taken from \citet{miloneetal2023}, or, for clusters not studied in this
paper from available literature.  We take the lifetimes of the clusters from \citet{baumgardtetal2013}, who determined lifetimes assuming that the clusters move in 
circular cluster orbits around the center
of their parent galaxy. This seems to be a good approximation at least for the star clusters of the LMC \citep{bennetetal2022}.
Table~\ref{table:lmcsmc} gives the parameters adopted in the fitting of the cluster CMDs together with the derived mass function slopes.
\begin{table*}
\caption{Details of the performed $N$-body simulations of star clusters evolving in tidal fields. Columns 3 and 4 contain the initial cluster mass and half-mass radius. Column 5 contains the assumed 
black hole retention fraction. Columns 6 and 7 contain the semi-major axis and eccentricity of the cluster orbits. Column 8 contains the remaining mass fraction at $T=13.5$ Gyr and column 9 the 
estimated lifetimes of the clusters.}
\begin{center}
\begin{tabularx}{0.75\textwidth}{crcccrccc}
\hline
\multirow{2}{*}{Model} &  \multirow{2}{*}{$N_{Stars}$} & $M_{Ini}$ &  $r_{h,Ini}$ & \multirow{2}{*}{$f_{BH}$} & $a_{Orb}$ & \multirow{2}{*}{$e_{Orb}$} & \multirow{2}{*}{$M_{Fin}/M_{Ini}$} & $T_{Diss}$ \\
 & & [M$_\odot$] &  [pc] & & [pc] &  & &  [Gyr] \\
\hline
 1 &  70,000 & $7.0 \cdot 10^5$  & 3.0 & 0.10  & 10000 & 0.00 & 0.197  & 21.7 \\
 2 &  70,000 & $7.0 \cdot 10^5$  & 3.0 & 0.10  &  5000 & 0.00 & 0.000  & 12.9 \\
 3 &  70,000 & $7.0 \cdot 10^5$  & 3.0 & 0.10  &  7500 & 0.00 & 0.104  & 16.4 \\
 4 &  70,000 & $7.0 \cdot 10^5$  & 2.0 & 0.10  &  6500 & 0.50 & 0.000  & 12.4 \\
 5 &  70,000 & $7.0 \cdot 10^5$  & 3.0 & 0.10  & 15000 & 0.00 & 0.294  & 34.4 \\
 6 &  70,000 & $7.0 \cdot 10^5$  & 5.0 & 0.10  & 15000 & 0.00 & 0.231  & 26.1 \\
 7 &  70,000 & $7.0 \cdot 10^5$  & 3.0 & 0.10  & 20000 & 0.00 & 0.340  & 48.9 \\
 8 &  70,000 & $7.0 \cdot 10^5$  & 5.0 & 0.10  & 20000 & 0.00 & 0.296  & 37.3 \\
 9 & 131,000 & $1.3 \cdot 10^5$  & 3.0 & 0.10  &  5000 & 0.00 & 0.159  & 19.0 \\
10 & 131,000 & $1.3 \cdot 10^5$  & 2.0 & 0.10  &  6500 & 0.50 & 0.087  & 15.7 \\
11 & 131,000 & $1.3 \cdot 10^5$  & 2.0 & 0.10  &  6300 & 0.60 & 0.044  & 14.4 \\
12 & 131,000 & $1.3 \cdot 10^5$  & 3.0 & 0.10  & 10000 & 0.00 & 0.296  & 36.4 \\
13 & 131,000 & $1.3 \cdot 10^5$  & 5.0 & 0.10  & 14000 & 0.44 & 0.219  & 24.7 \\
14 & 131,000 & $1.3 \cdot 10^5$  & 3.0 & 0.10  &  3000 & 0.00 & 0.000  & 13.1 \\
15 & 131,000 & $1.3 \cdot 10^5$  & 5.0 & 0.10  &  7500 & 0.00 & 0.054  & 14.1 \\
16 & 131,000 & $1.3 \cdot 10^5$  & 5.0 & 0.10  & 10000 & 0.00 & 0.285  & 30.3 \\
17 & 131,000 & $1.3 \cdot 10^5$  & 5.0 & 0.10  & 11250 & 0.33 & 0.245  & 26.5 \\
18 & 131,000 & $1.3 \cdot 10^5$  & 5.0 & 0.10  & 20000 & 0.00 & 0.277  & 29.7 \\
19 & 200,000 & $2.0 \cdot 10^5$  & 3.0 & 0.10  &  3000 & 0.00 & 0.021  & 13.8 \\
20 & 300,000 & $3.0 \cdot 10^5$  & 3.0 & 0.10  &  3000 & 0.00 & 0.177  & 19.4 \\
\hline
\end{tabularx}
\end{center}
\label{tab:nbody}
\end{table*}

The lifetimes and relaxation times of the LMC/SMC clusters are generally larger than their
cluster ages, hence we do not expect the stellar mass function of these clusters to have significantly changed since their formation.
Fig.~\ref{fig:lmcsmcmf} shows the derived mass function slopes for the LMC/SMC star clusters. The left panel depicts the slopes derived for the full range of stellar
masses that we can fit. It can be seen that there is a gradual change of the stellar mass function around 1 M$_\odot$. High-mass stars with masses
$m>1$ M$_\odot$ have slopes close to a Salpeter mass function ($\alpha=-2.3$) while for low-mass stars the average power-law slope is around $\alpha \approx -1.4$. The transition between both slopes happens at a mass
of about $m=1$ M$_\odot$. This conclusion is strengthened by the right panel in Fig.~\ref{fig:lmcsmcmf}, where we have determined mass functions separately
for stars with masses below and above 1 M$_\odot$. We have restricted the fits in this panel to clusters where either the low-mass limit is below 0.60 M$_\odot$ or
the high-mass limit is above 1.60 M$_\odot$. It can be seen that the derived slopes split into two well separated groups. The high-mass star function is compatible with
a Salpeter slope with $\alpha=-2.3$ in essentially all clusters, while the low-mass stellar mass function has an average slope of $\alpha \approx -1.4$, compatible
with what we found for Galactic globular clusters.

Combining the Milky Way and LMC/SMC results, and assuming that the Milky Way GCs had a high-mass star mass function similar to the LMC/SMC clusters, we can therefore describe
the initial mass function of the clusters in our sample as a three-stage power-law with:
\begin{eqnarray}
\nonumber \alpha_{High} & = & -2.30 \pm 0.15 \;\;\; \mbox{for} \;\; m > 1 \; M_\odot \\
\nonumber \alpha_{Med} & = & -1.65 \pm 0.20 \;\;\; \mbox{for} \;\; 0.4 \; M_\odot < m < 1 \; M_\odot \\
\nonumber \alpha_{Low} & = & -0.30 \pm 0.20 \;\;\; \mbox{for} \;\; m<0.4 \; M_\odot
\end{eqnarray}

Alternatively, the initial mass function can also be described as a log-normal mass function with $\log M_C=-0.36$ and $\sigma_C=0.28$ for stars with masses $m<1$ M$_\odot$
followed by a power-law mass functions for higher mass stars with a slope $\alpha = -2.3$.

\subsection{$N$-body simulations of star cluster evolution}

Having determined the initial cluster mass function from the dynamically least evolved clusters, we now turn our attention to fitting the full cluster sample.  In order to do this,
we first ran $N$-body simulations of star clusters dissolving in tidal fields
starting from the three-stage power-law mass function determined in sec.~\ref{sec:imf}. We will use these simulations in the next sections to determine the mass functions of evolved star clusters
as well as to derive constraints on the black hole retention fraction of star clusters.
\begin{table*}
\caption{Power-law mass function slopes $N(m) \sim m^{\alpha}$ and mass limits used as input values for our grid of $N$-body simulations.}
\begin{tabularx}{\textwidth}{cccccccccccccc}
\hline
\multirow{2}{*}{Model} & \multirow{2}{*}{$M(t)/M_0$} & $m_{Low}$ & $m_{Up}$ & \multirow{2}{*}{$\alpha$} & $m_{Low}$ & $m_{Up}$ & \multirow{2}{*}{$\alpha$} & $m_{Low}$ & $m_{Up}$ & \multirow{2}{*}{$\alpha$} & $m_{Low}$ & $m_{Up}$ & \multirow{2}{*}{$\alpha$} \\
 &  & [M$_\odot$] & [M$_\odot$] & & [M$_\odot$] & [M$_\odot$] & & [M$_\odot$] & [M$_\odot$] & & [M$_\odot$] & [M$_\odot$] & \\
\hline
1 & 1.00 & 0.10 & 0.40 & -0.35 & 0.40 & 1.00 & -1.65 & 1.00 & 6.50 & -2.30 & 6.50 & 100.0 & -2.30 \\
2 & 0.30 & 0.10 & 0.40 & -0.20 & 0.40 & 1.00 & -1.35 & 1.00 & 6.50 & -2.25 & 6.50 & 100.0 & -2.80 \\
3 & 0.22 & 0.10 & 0.40 & -0.05 & 0.40 & 1.00 & -1.05 & 1.00 & 6.50 & -2.25 & 6.50 & 100.0 & -3.15 \\
4 & 0.15 & 0.10 & 0.40 & $\,$0.25 & 0.40 & 1.00 & -0.65 & 1.00 & 6.50 & -2.20 & 6.50 & 100.0 & -3.20 \\
5 & 0.10 & 0.10 & 0.40 & $\,$0.50 & 0.40 & 1.00 & -0.05 & 1.00 & 6.50 & -1.90 & 6.50 & 100.0 & -3.20 \\
6 & 0.05 & 0.10 & 0.40 & $\,$0.80 & 0.40 & 1.00 &  0.30 & 1.00 & 6.50 & -1.60 & 6.50 & 100.0 & -3.80 \\
\hline
\end{tabularx}
\label{tab:mfrange}
\end{table*}

In total we ran 20 $N$-body simulations of star clusters using NBODY7 \citep{nitadoriaarseth2012}. The clusters
contained between $70,000$ to $300,000$ stars initially and moved in either circular or elliptic orbits through an isothermal galaxy with a constant
circular velocity of $v_c=240$ km/sec. The clusters followed \cite{king1966} density profiles with dimensionless concentration parameter $c=1.5$ initially.
We ran the simulations with an assumed neutron star and black hole retention fraction of 10\%, i.e. 90\% of the formed black holes
and neutron stars were given large velocity kicks upon their formation so that they left their parent clusters.  The
retention fraction was applied to every formed neutron star and black hole independent of its mass. We
took snapshots spaced by 500 Myr during the simulations and use the snapshots between 8 and 13.5 Gyr to determine the mass function of the remaining stars.
By 13.5 Gyr, the studied clusters had lost between 20\% to 100\% of their initial stars. Table~\ref{tab:nbody} gives details of the
performed $N$-body simulations. 
\begin{figure}
\begin{center}
\includegraphics[width=0.95\columnwidth]{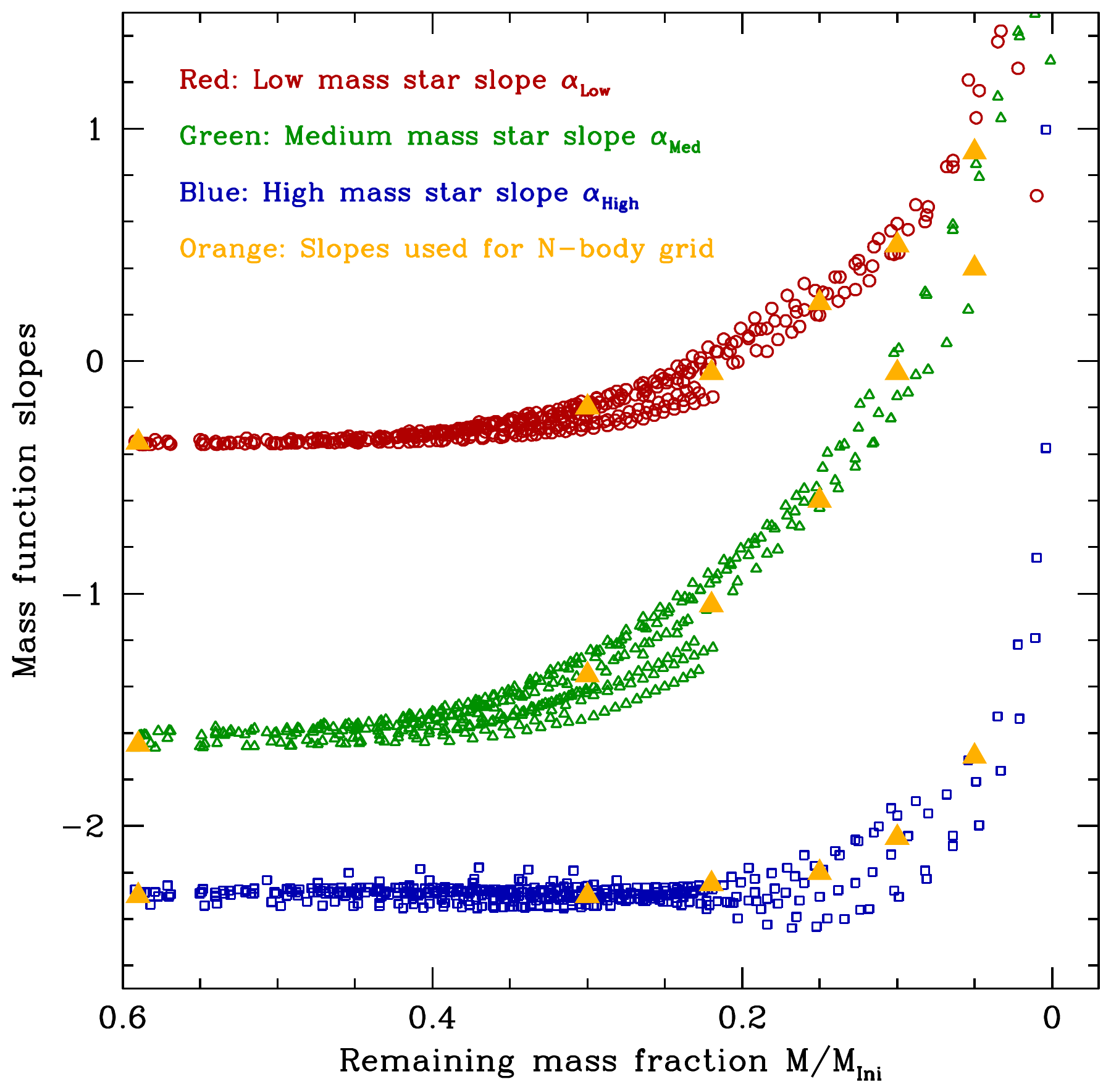}
\end{center}
\vspace*{-0.5cm}
\caption{Evolution of the mass function slopes $\alpha$ of low (open circles), intermediate-mass (green triangles) and high-mass (blue squares) stars in the $N$-body simulations
as a function of the remaining mass fraction $M/M_{Ini}$. For all three mass groups the mass functions evolve towards positive values due to mass segregation and the
preferential loss of lower mass stars. The evolution is stronger towards the final stages of cluster evolution and for the intermediate-mass stars. Orange triangles mark the
mass function values that are given in Table~2 and that are used for the grid of $N$-body simulations described in sec.~\ref{sec:grid}.}
\label{fig:mfevo}
\end{figure}

Fig.~\ref{fig:mfevo} depicts the change in the mass function slopes for low, intermediate-mass and high mass stars in the $N$-body simulations
as a function of the mass lost in the clusters. We depict only the evolution starting from the point when the clusters contain 60\% of their initial mass since the
mass lost up to that point is mainly due to stellar evolution within the first Gyr of evolution. When deriving the mass function slopes, we used for all stars their initial masses 
in order to remove the effect of stellar evolution of massive stars on the mass function. For the same reason we fit only the mass range from
1.0 to 8.0 M$_\odot$ for the high-mass stars since most of the more massive stars were removed by natal velocity kicks, creating a discontinuity in the
mass function. It can be seen that the clusters become increasingly depleted in low-mass stars
as time progresses. The evolution is particularly strong for intermediate-mass stars (0.40 M$_\odot < m < 1.0$ M$_\odot$) and towards the later stages of evolution. The slower
mass function evolution in the beginning is most likely due to the fact that clusters first need to become mass segregated before significant changes occur to their internal 
mass functions. Overall the mass function evolution proceeds in a very similar way in the different clusters despite the fact that the simulations span
a wide parameter range. Hence it can be expected that the evolution of real star clusters proceeds along a similar path.

The orange triangles in Fig.~\ref{fig:mfevo} mark the points when we determine (averaged) mass function slopes from the $N$-body simulations. We list the derived mass
function values in Table~\ref{tab:mfrange}. From the initial masses of the remaining neutron stars and black holes we also derive an additional slope for the 
highest mass stars. The derived values are somewhat uncertain due to the small number of remaining stars in the clusters. Nevertheless they show the strong depletion
of the more massive black holes through dynamical encounters and binary formation in the cluster cores, which leads to a strong steepening of the mass function of
the remaining black holes.
\begin{figure*}
\begin{center}
\includegraphics[width=0.97\textwidth]{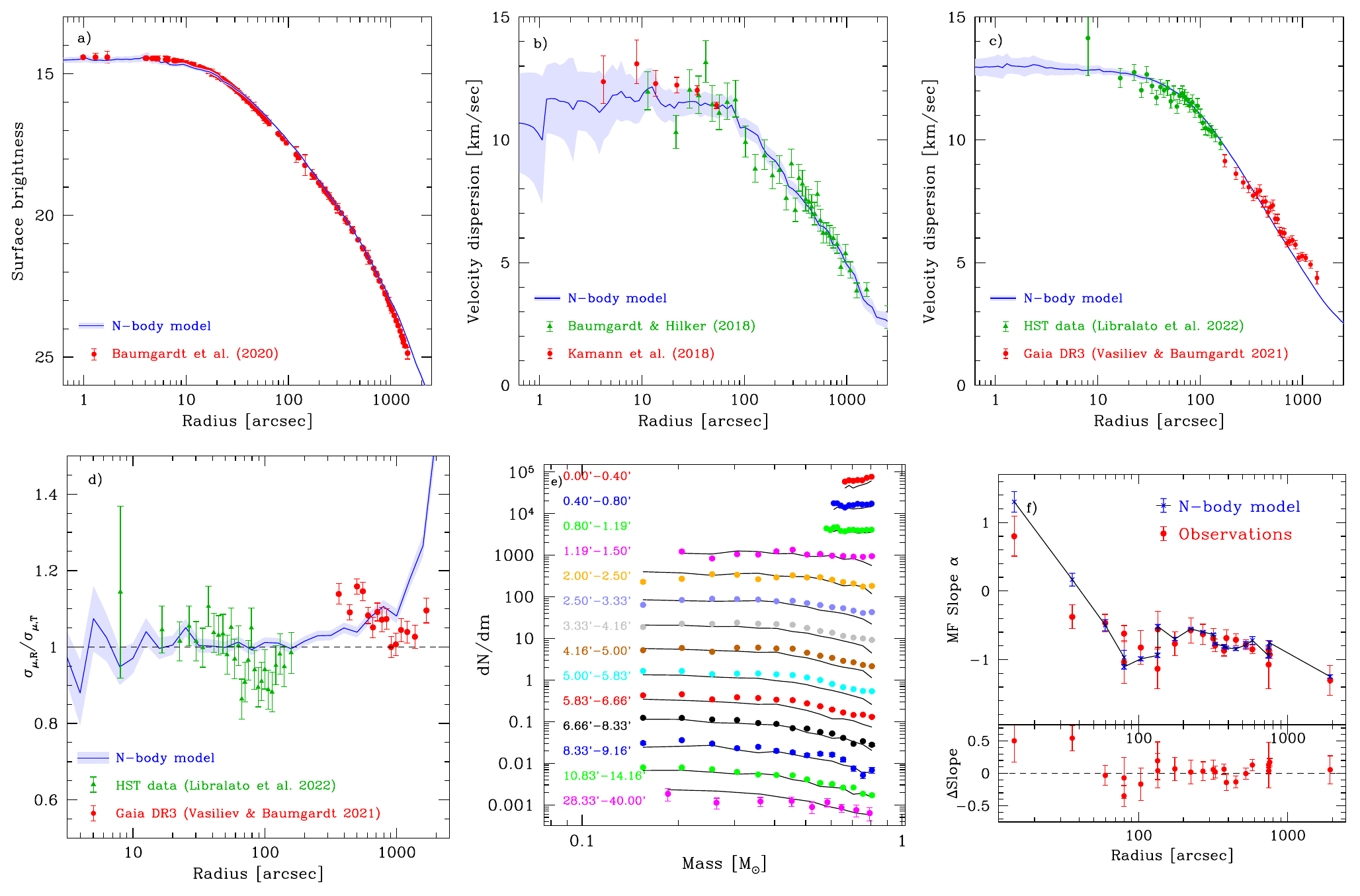}
\end{center}
\vspace*{-0.5cm}
\caption{Comparison of the best-fitting $N$-body model to the observed surface density profile (panel a), line-of-sight velocity dispersion profile (panel b), proper motion velocity 
 dispersion profile (panel c) velocity anisotropy profile (panel d), individual stellar mass functions at different radii (panel e) and the slope of the mass function as a function 
 of radius (panel f) for the globular cluster
NGC~104. It can be seen that the $N$-body model is in general in good agreement with the observations.}
\label{fig:ngc104}
\end{figure*}

\subsection{Derivation of the global mass function}
\label{sec:grid}

The derivation of the global mass functions for clusters that are dynamically evolved and have incomplete spatial coverage from the HST photometry
was done similar to \citet{baumgardt2017} and \citet{baumgardthilker2018} by fitting a large grid of $N$-body models to each observed cluster and 
finding the model that best fits all available data for each cluster
We give a brief summary of this fitting procedure below, more details can be found in \citet{baumgardt2017} and \citet{baumgardthilker2018}. 
\begin{figure*}
\begin{center}
\includegraphics[width=0.85\textwidth]{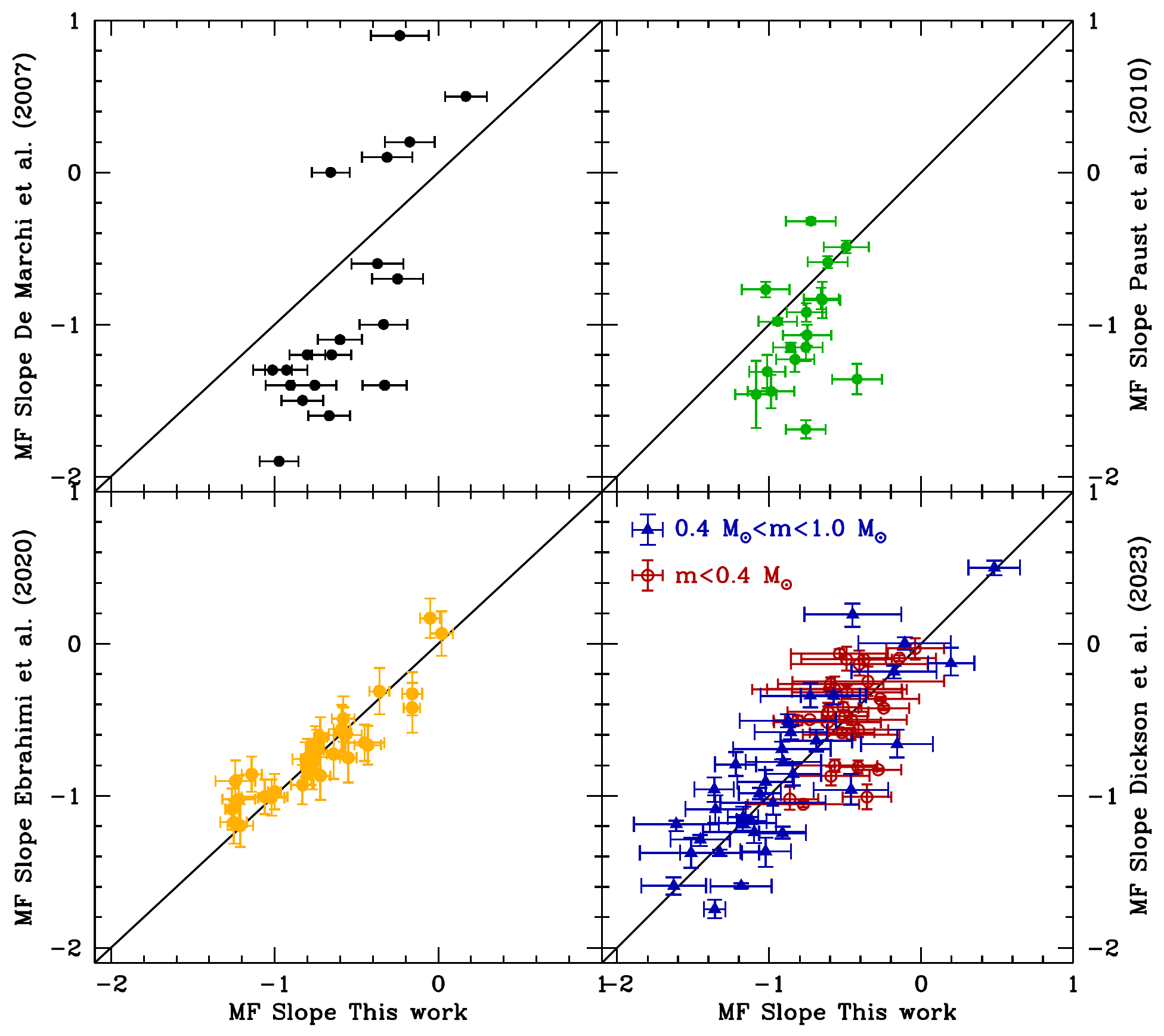}
\end{center}
\vspace*{-0.5cm}
\caption{Comparison of the mass function slopes derived in this work against those derived by \citet{demarchietal2007} (top left panel), \citet{paustetal2010} (top right panel),
 \citet{ebrahimietal2020} (bottom left panel) and \citet{dicksonetal2023} (bottom right panel) for the clusters in common. We use single power-law slopes for the comparison
  with the literature papers except when comparing with \citet{dicksonetal2023}, where we fit mass functions to stars in the low and intermediate mass regime separately.}
\label{fig:compmf}
\end{figure*}

We first used the six sets of mass function values listed in Tab.~\ref{tab:mfrange} and ran a large grid of $N$-body simulations using these values.
For each mass function, we used eight values for the initial half-mass radius $r_h$, spaced roughly evenly in $\log r_h$ between 2~pc and 35~pc
and six values for the initial concentration index $c$ of the initial \citet{king1962} density profile between $c=0.2$ to $c=2.5$. All $N$-body
simulations were isolated simulations having $10^5$ stars initially and were run for 13.5 Gyr. Like in the simulations in the preceding section,
we assumed a 10\% retention fraction of black holes and neutron stars in these models.

The fitting of the Galactic globular clusters and LMC/SMC star clusters was then done by selecting the snapshot closest in time to the age of an observed cluster from each $N$-body simulation 
and by then fitting these snapshots to the observational data available for each cluster. We scaled each $N$-body model along lines of constant relaxation time $T_{RH}$ to the same half-light radius of
an observed cluster, using the cluster distances that were derived by \citet{baumgardtvasiliev2021} by combining a variety
of individual distance determinations. We then calculated the velocity dispersion and surface density profiles for each $N$-body model from the distribution of
its bright stars. We also calculated a sky projection of each $N$-body model centered around the position of each observed cluster and selected
stars from the $N$-body models that are in the same region of
the sky as the observed HST fields. We then derived stellar mass function slopes for these regions and for the same radial annuli for which we have observational
data.

We then interpolated in our grid of models and derived the model that provides the best fit to the observed surface density profile from \citet{baumgardtetal2020},
the observed velocity dispersion profile, and to the observed mass function slopes at different radii that we calculated in this paper. We determine the best-fitting $N$-body model 
through $\chi^2$ minimization against the observed data and adopt as best-fitting cluster parameters the parameters of this model. From the models with $\chi^2 < \chi_{min}^2+1$ we 
also obtain error bars on the cluster parameters. In order to reflect the influence that e.g. uncertainties in the cluster ages have on the derived mass 
functions slopes, we add an uncertainty of $\Delta \alpha = 0.10$ in quadrature to the global mass function errors derived this way and adopt the resulting
value as the final error.

In order to improve the accuracy with which we can reproduce the cluster parameters, we added kinematic data published in
recent years to the kinematic data from \citet{baumgardthilker2018}. Most of this data comes from large scale radial velocity surveys targeting Milky Way stars
like Gaia DR3 \citep{gaiadr3main,gaiadr3radvel}, Apogee DR17 \citep{apogeedr17main}, Lamost DR7 \citep{cuietal2012}, Galah DR3 \citep{buderetal2021}
and the WAGGS survey \citep{dalgleishetal2020}. We list additional sources
improving kinematic data for particular clusters in the Appendix.
From our fits we not only derived the stellar mass functions but also a range of other cluster parameters including
total masses, core and half-mass radii and relaxation times $T_{RH}$, which we will use further below. 
Fig.~\ref{fig:ngc104} shows as an example of the fitting procedure our fit for the cluster NGC~104. It can be seen that our best-fitting $N$-body model 
reproduces the surface density and velocity dispersion profiles as well as the individual mass functions at various radii fairly well, despite the fact that
our models need to fit a large amount of observational data with only few free parameters. 
We make the full set of comparisons as well as the derived parameters available on a dedicated
website\footnote{https://people.smp.uq.edu.au/HolgerBaumgardt/globular/}. Finally, we also calculated the dissolution time $T_{Diss}$ for each cluster 
using the approach described in sec.~3.2 of \citet{baumgardtetal2019a}.
\begin{figure*}
\begin{center}
\includegraphics[width=0.99\textwidth]{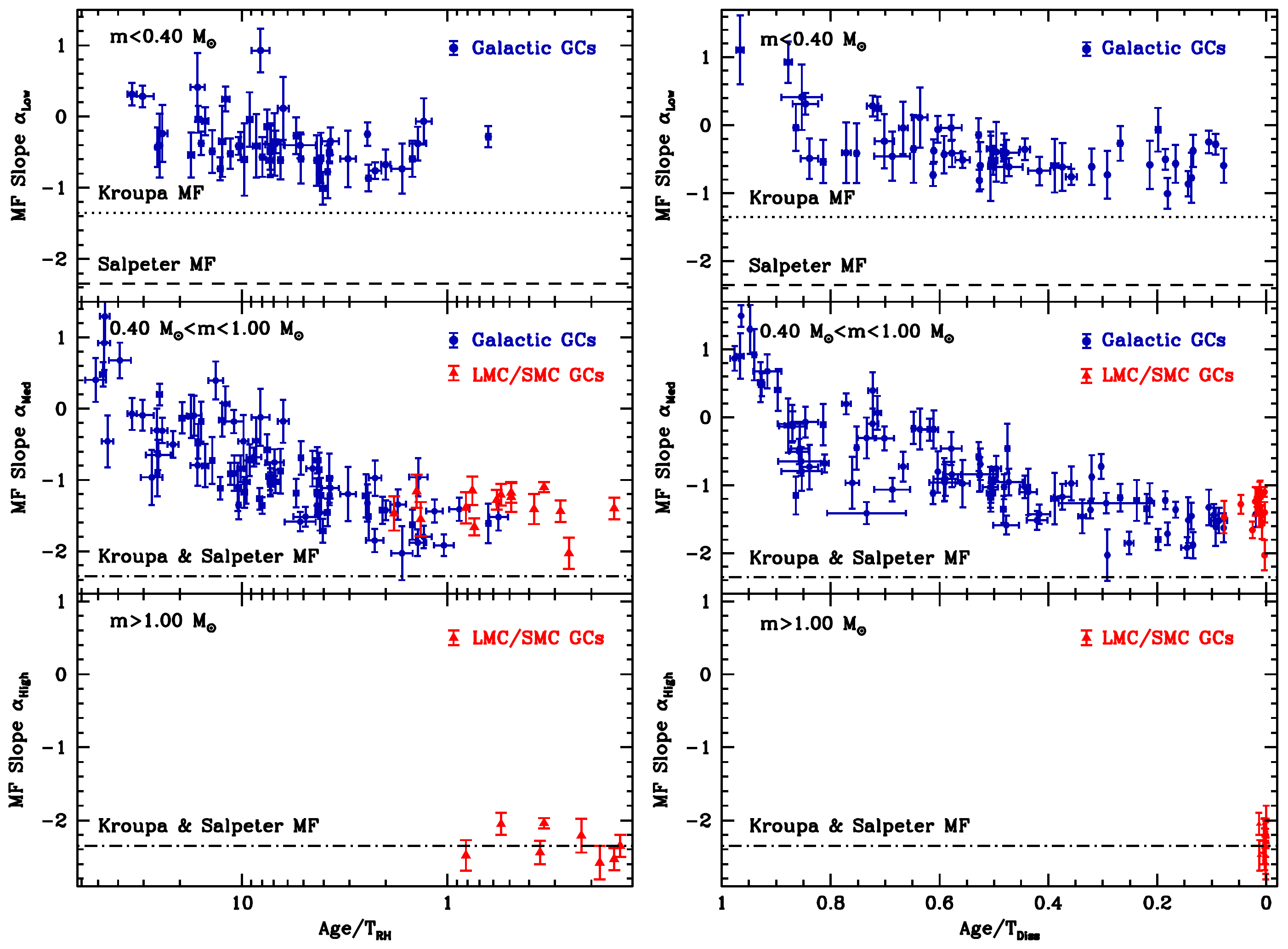}
\end{center}
\vspace*{-0.5cm}
\caption{Global mass function slopes $\alpha$ as a function of the dynamical ages of globular clusters (left panel) and the ratio of their
ages to their lifetimes (right panel). Globular clusters follow a narrow sequence going from clusters with negative mass function slopes
(many low-mass stars) and relaxation times/lifetimes that are larger than their ages to clusters highly depleted in low-mass stars for clusters
with short relaxation times/lifetimes. Clusters are also considerably more depleted in low-mass stars than canonical Kroupa/ Salpeter mass functions.}
\label{fig:mfres}
\end{figure*}

We present the derived mass function slopes in Tables~\ref{table:main} and \ref{table:lmcsmc}. In these tables we have characterised the mass functions by both single power-law
mass functions $N(m) \sim m^{\alpha}$ over the whole observed mass range, as well as by two-stage power-law mass functions for clusters that have a sufficiently wide
mass coverage to allow low-mass/high-mass stars to be fitted separately. For the two-stage power-law mass functions we assumed fixed break masses of 0.4 M$_\odot$ 
(Milky Way GCs) and 1.0 M$_\odot$ (LMC/SMC clusters).
In general two-stage power-law mass functions provide significantly better fits and we will therefore mostly use these in the rest of the paper.

Fig.~\ref{fig:compmf} compares the mass function slopes that we find in this work with mass function determinations of Galactic globular clusters
by \citet{demarchietal2007}, \citet{paustetal2010}, \citet{ebrahimietal2020} and \citet{dicksonetal2023} for the clusters in common. The first three papers characterised
the stellar mass function by a single power-law slope $N(m) \sim m^{\alpha}$ over the whole range of masses fitted. We therefore also use our single power-law mass
function fits for comparison. It can be seen that the mass function slopes determined by \citet{demarchietal2007} and \citet{paustetal2010} span a larger 
range in values compared to our data and are also lower on average. This could at least partly be due to the fact that their lower-mass limits are usually higher, causing 
their mass functions to be 
more strongly dominated by stars with masses $m>0.4$ M$_\odot$ which have a steeper mass function slope. Some of the mass function slopes quoted by \citet{demarchietal2007} are also
locally measured ones that were not corrected for mass segregation. This will lead to an under or over estimation of the global mass function slopes if the 
local mass function was determined in the cluster centre or the cluster halo. 
We obtain excellent agreement with the mass functions derived by \citet{ebrahimietal2020} with almost all clusters being in agreement within the quoted error bars.
This is despite the fact that the fitted isochrones as well as the method how the mass functions were derived are different between our paper and \citet{ebrahimietal2020}. The bottom right panel
finally compares our mass function slopes with the ones from \citet{dicksonetal2023}. Here we compare the mass function slopes separately for stars with masses above and below
0.4 M$_\odot$.  The slopes from \citet{dicksonetal2023} are not completely independent of ours since they also used our HST star counts. However the method
to get from the star count data in different HST fields to the global mass function is different to ours. Nevertheless, the agreement between the derived 
mass function slopes is very good both for stars with $m>0.4$ M$_\odot$ and for stars with $m<0.4$ M$_\odot$.

\subsection{Globular cluster mass functions}

We next discuss the mass function of the full sample of MW globular clusters and LMC/SMC stars clusters, derived by fitting the grid of $N$-body simulations calculated
in the previous section against their observed mass function slopes.
Fig.~\ref{fig:mfres} shows the global mass function slopes of Galactic globular clusters and LMC/SMC star clusters as a function of 
the dynamical age (left panel) and the fractional lifetime of the clusters (right panel). We have defined the dynamical age as the ratio of the age
of a cluster over its relaxation time and the fractional lifetime as the ratio of the cluster age to the estimated dissolution time. We fit the
mass function of each cluster using either only the low mass stars $m<0.40$ M$_\odot$ (upper panels), the intermediate-mass stars with 0.40 M$_\odot< m<1.0$ M$_\odot$ (middle panels) or
high mass stars with $m>1$ M$_\odot$ (lower panels). Since Galactic globular clusters
do not contain high-mass main-sequence stars with $m>1$~M$_\odot$, we have results for them only for the upper and middle panels, while for the LMC/SMC clusters we have no results
for the $m<0.40$ M$_\odot$ stars since these are too faint to be observed due to the large cluster distances.

It can be seen that we obtain a strong correlation between the mass function slopes and either the dynamical age or
fractional dissolution time for the low and intermediate mass stars, with Spearman-rank order coefficients equal to $r=0.85$ and $r=0.81$ respectively. 
In particular, the mass function slopes for low and intermediate mass stars are more strongly negative for the dynamically least evolved clusters and
become gradually less negative for clusters with smaller relaxation times and smaller lifetimes (relative to their ages).
As will be discussed further below, the correlation of mass function slopes
with dynamical ages and fractional lifetimes is likely due to the internal evolution of the clusters and not a sign of initial variations. If this 
is the case, the present-day mass functions of clusters on the right hand side in both diagrams should reflect their initial mass functions, justifying the
cluster selection made in sec.~4.1. There is also very good agreement between the mass functions of the Galactic globular clusters and the LMC/SMC clusters 
in the regions of overlap. For the high mass stars we have data only for the LMC/SMC star clusters. Since these clusters are on average more extended than Galactic globular
clusters, they have large relaxation times and our data does not allow us to see an evolution of the high-mass star mass function as we can only probe the less 
evolved and presumably primordial distribution. The same applies to the fractional lifetime distribution, due to the young ages of the clusters and their long lifetimes times, 
the Age/$T_{Diss}$ ratio is less than 0.1 for all LMC/SMC clusters.
\begin{figure*}
\begin{center}
\includegraphics[width=0.95\textwidth]{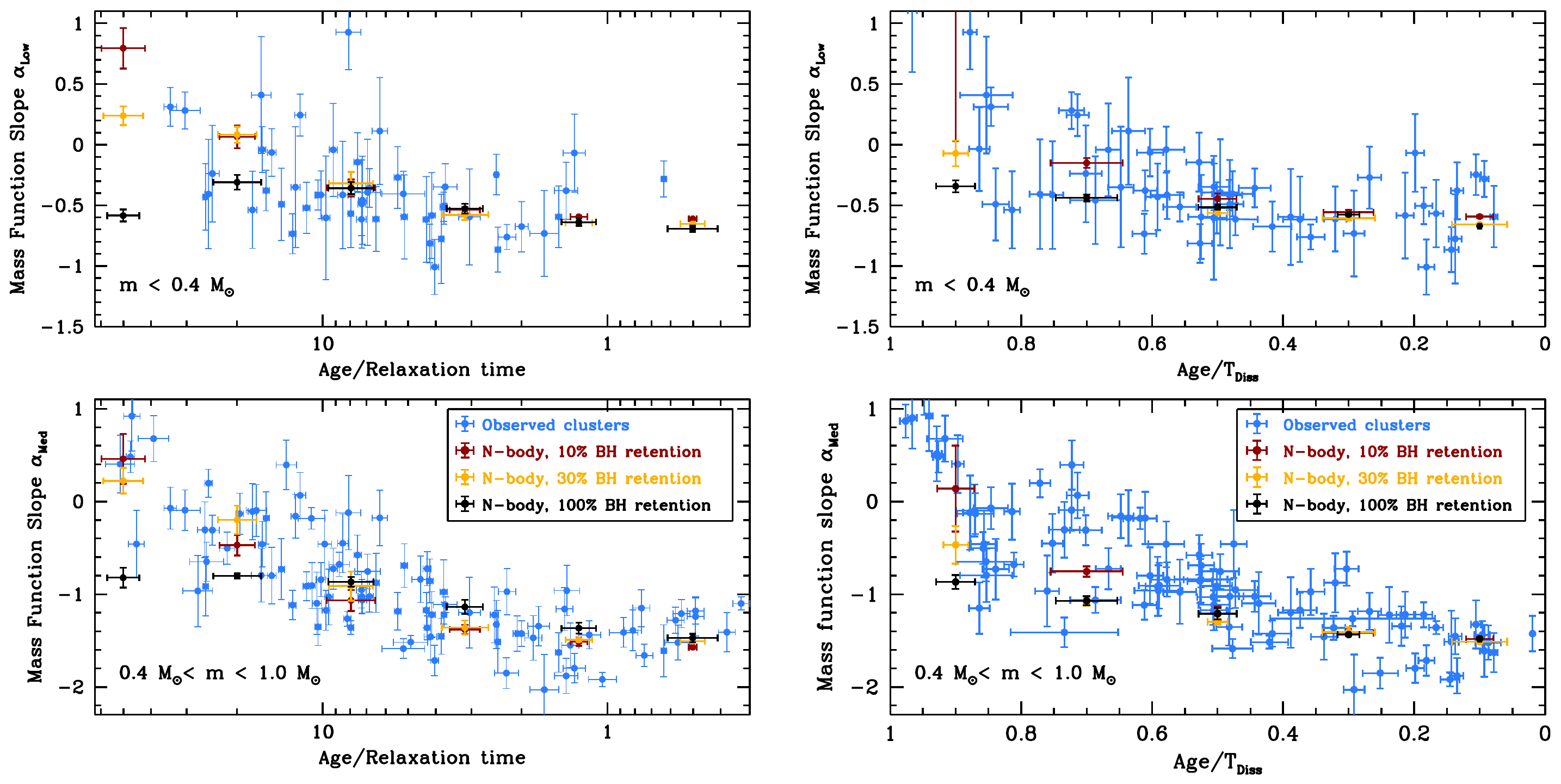}
\end{center}
\vspace*{-0.2cm}
\caption{Comparison of the derived mass function slopes for Galactic GCs (blue circles) with the results of $N$-body simulations of star clusters dissolving in tidal fields
with an assumed BH retention fraction (upon formation) of 10\% (red), 30\% (orange) and 100\% (black). $N$-body models with 10\%  BH retention fraction follow the observed trends with dynamical age
and fractional lifetime very well. 30\% BH retention models reproduce the tend with dynamical age well, but reproduce the trend with fractional lifetime less well.
100\% BH retention models can't reproduce these trends since they are not able to produce clusters strongly depleted in low-mass stars.}
\label{fig:cmpnbody}
\end{figure*}

Using only the least evolved clusters with relaxation times and lifetimes larger than a Hubble time, we find average mass function slopes of $\alpha_{Low} \approx -0.3$ for $m<0.4$ M$_\odot$ 
stars, $\alpha_{Low} \approx -1.6$ for stars with
0.4 M$_\odot<m<1.0$ M$_\odot$ and -2.3 for stars with $m>1.0$ M$_\odot$. A Kroupa (2001) mass function has a slope of $\alpha=-2.3$ for stars more massive than 0.5 M$_\odot$ 
and a slope of $\alpha=-1.3$ for less massive stars, while a Salpeter mass function has a slope of $\alpha=-2.3$ for all stars (see dashed and dotted lines in Fig.~\ref{fig:mfres}).  
Our results from the expanded cluster sample again argue for a low-mass star function in globular clusters with significantly fewer low-mass stars compared to a Kroupa/Salpeter 
mass function. Our mass function below 0.8 M$_\odot$ is in good agreement with the power-law mass function slope of $-1.3$ that \citet{leighetal2012} found from
comparing the inner mass functions 27 globular clusters with the results of Monte Carlo simulations. We also confirm earlier results by \citet{cadelanoetal2020b} for the global mass functions of NGC~7078 and NGC~7099 as well as
\citet{henaultbrunetetal2020} for the mass function of NGC~104.

\subsection{Comparison with $N$-body simulations}
\label{sec:compnbody}

We next investigate if the observed correlations of the stellar mass function with dynamical age and fractional lifetime that we find for Milky Way globular clusters
can be explained by the dynamical cluster evolution and the constraints the results can put on the black hole retention fraction in star clusters.
We again use the $N$-body simulations from Table~1 for 10\% retention fraction. In order to test the dependence of the results on the assumed black hole retention fraction,
we also re-run all simulations in Table~1 with 30\% and 100\% BH retention fractions. In these new simulations, we keep the neutron star retention fraction at 10\%.
\begin{figure*}
\begin{center}
\includegraphics[width=0.90\textwidth]{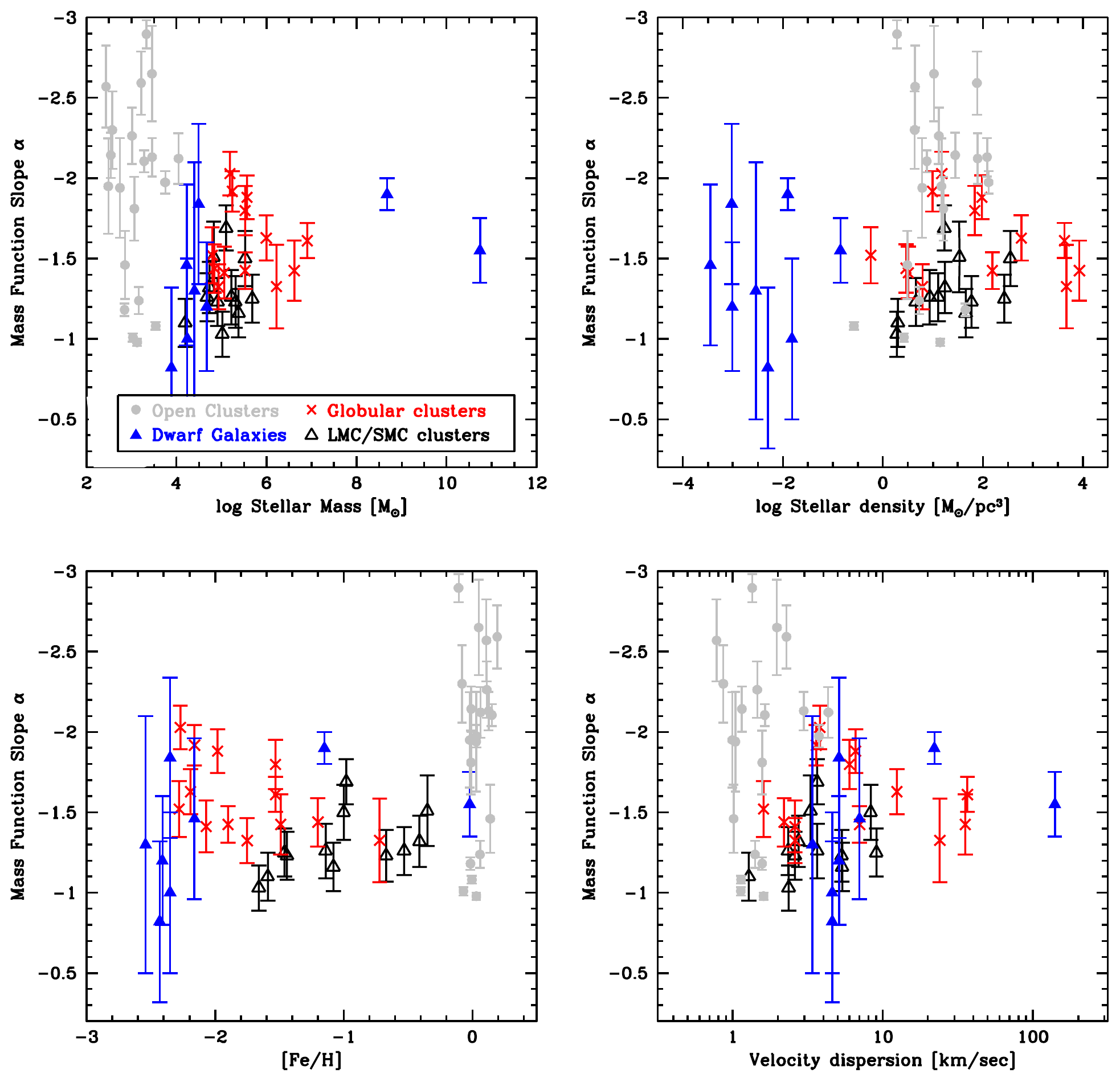}
\end{center}
\vspace*{-0.2cm}
\caption{Stellar mass function slopes as a function of (clockwise from top left) the mass of the stellar system, its stellar density, velocity dispersion and metallicity.
Star clusters studied in this paper are shown by red crosses and black, open triangles. We only show clusters with relaxation times and lifetimes larger than their ages 
for which the present-day mass function should still be close to the initial one. In addition we also show stellar mass functions for Milky Way open clusters from
\citet{ebrahimietal2022} and \citet{cordonietal2023} (grey circles) as well
as those for field stars in several nearby dwarf galaxies (blue filled triangles). There does not seem to be a correlation of the mass function slope with any
of the depicted parameters or stellar systems. The only exception seem to be open cluster mass functions, which are on average more bottom-heavy.}
\label{fig:cmdmfall}
\end{figure*}

Fig.~\ref{fig:cmpnbody} shows the comparison of the mass function slopes of observed globular clusters with the results of the $N$-body simulations. It can be seen that
models with a 10\% retention fraction of black holes reproduce the observed mass function trends with dynamical age and fractional lifetime very closely, showing that
the differences in the stellar mass function between different clusters are unlikely to arise due to initial differences, but can be explained by dynamical changes due to
mass segregation and the preferential loss of low-mass stars. The models also show that the slope of the low-mass star mass function changes significantly less
over the course of evolution compared to the slope of stars in the mass range 0.4 to 1.0 M$_\odot$. The reason for this behavior is probably that low mass stars are
pushed towards the outer parts where the relaxation time is long and there is little further mass segregation between these stars. Hence they are lost at a similar rate
independent of their mass.

Clusters starting with a 100\% black hole
retention fraction never reach a state where they are highly depleted in low-mass stars. This is because a large number of stellar-mass black holes
in a star cluster prevents mass segregation between the lower mass stars \citep[e.g.][]{lutzgendorfetal2013,weatherfordetal2018}. As a result, clusters with many black holes
are unable to reproduce the observed mass function evolution, especially for the dynamically more evolved clusters. Clusters with 30\% BH retention rates can reproduce 
the trend with dynamical age but don't reproduce the trend with dissolution time towards the final stages. We therefore conclude that a high initial BH retention fraction 
is ruled out by our data. Given our chosen mass function, and assuming a 30\% retention fraction of black holes leads to a typical ratio in the number of black holes
to the total number of stars of around $N_{BH}/N_* \approx 1.0 \cdot 10^{-3}$ directly after BH formation. This ratio further decreases due to BH binary formation and subsequent hardening
and ejections of the black holes from the cluster centres. Assuming a decrease by a factor of two to five in the number of BHs surviving up to a Hubble time, we predict between 30
to 100 (300 to 1000) remaining BHs
in a $10^5$ M$_\odot$ ($10^6$ M$_\odot$) globular cluster. These estimates are in good agreement with estimates for the number of stellar-mass BHs inferred from observations
of the surface density profiles \citep{arcaseddaetal2018,askaretal2018} and the internal amount of mass segregation \citep{weatherfordetal2018} of Galactic globular clusters.
They are also in agreement with what \citet{dicksonetal2023} find from a comparison of multi-mass King models with the observed kinematics of globular clusters.

\subsection{Environmental dependency of the stellar mass function}

In order to explore the influence of the external environment on the stellar mass function, we depict in Fig.~\ref{fig:cmdmfall} the mass-function slopes that we have derived as a function of
different cluster parameters. Shown are (clockwise from top-left), the cluster mass, the average density inside the half-mass radius of the cluster, the central velocity
dispersion of the cluster and the cluster metallicity. We restrict our fits to intermediate-mass stars in the range 0.4 M$_\odot < m < 1.0$ M$_\odot$ since this is the mass function
range for which we have the most data and the only one where LMC/SMC clusters and globular clusters overlap. We depict only clusters with remaining lifetimes several times larger
than their ages since for these the
present-day mass functions should still be close to the initial ones. In order to correct for mass loss due to stellar evolution, we increase the present-day cluster
masses by a factor of two to get the initial masses. We also assume that stellar evolution induced mass loss leads to an adiabatic expansion of the cluster so that the initial
half-mass radius was half the present-day value. Both assumptions lead to a factor 16 increase of the initial density over the present-day one and a factor two increase of
the initial velocity dispersion. The true increase could be higher if the dynamical cluster evolution has led to an expansion of a cluster, however given the
large relaxation times of the depicted clusters, this is likely not a large effect.

We extend our sample beyond star clusters by adding the mass function determinations from \citet{gehaetal2013} and \citet{gennaroetal2018} for six ultra-faint dwarf galaxies
and from \citet{kaliraietal2013} for field stars in the SMC. We also add data about the IMF of Milky Way disc
stars in the solar neighborhood derived by \citet{sollima2019} based on Gaia DR2 parallaxes and magnitudes. These eight measurements form the galaxy sample in Fig.~\ref{fig:cmdmfall}.
We take the metallicities, masses, sizes and velocity dispersions of the dwarf galaxies from \citet{simon2019}. The galaxy data covers the sub-solar mass function, making it
directly comparable to the data from the cluster sample.
We finally add mass function determinations for open clusters in the solar neighborhood recently derived by \citet{ebrahimietal2022}  and \citet{cordonietal2023} based
on Gaia DR3 data. We use only clusters with relaxation times larger than their ages from both papers in order to minimise the influence of dynamical evolution and use the
mass function slopes for stars with $m<1$ M$_\odot$ to make the open cluster data comparable to the data of the other systems.

As can be seen, most mass functions are compatible with a slope of about $\alpha \approx -1.5$, without any obvious dependency 
of the stellar mass function slope on either the mass, the density, the metallicity or the internal velocity dispersion of the stellar system. The visual impression is
confirmed by Spearman rank order tests which do not indicate a significant correlation between any of the depicted parameters and the mass function slope.
\cite{markskroupa2010} suggested a systematic change of the initial stellar mass function with the density of the star forming cloud that forms a globular cluster. However, given our data, 
at least the sub-solar IMF seems more or less independent on density over nearly eight orders of magnitude. We furthermore see no evidence for a metallicity dependency
of the stellar IMF for globular clusters as has been argued by \citet{zonoozietal2016}. In addition, as discussed in sec.~4.4, stars with masses more massive than
1.0~M$_\odot$ seem to follow a Salpeter mass function for all studied LMC/SMC clusters, similar to the mass function seen for massive stars in a wide range of environments 
\citep{bastianetal2010}, again arguing against a variation of the high-mass IMF with metallicity.
The last conclusion is confirmed by \citet{dicksonetal2023}, who find that the kinematic data of globular clusters, when accounting for stars which have evolved into remnants
at the present-day, is fully compatible with a Salpeter IMF above 1~M$_\odot$. \citet{dicksonetal2023} also find no correlation between this mass function and cluster 
metallicity.

The only exception could be open clusters in the solar neighborhood, shown by grey circles in Fig.~\ref{fig:cmdmfall},
for which we find an average mass function slope of $\alpha=-1.9$. This is about 0.4 dex steeper than the average slopes that we find for the other systems
and could indicate a change in the IMF happening at either solar  metallicity, low masses, young ages, or a combination of these parameters. The open cluster data however also
shows a large scatter in the slopes for individual clusters, which range from about -1 to -3, and which seems to be much larger than what can be explained by errors in the data alone.  
Hence it is possible that there could still be some systematic error in the open cluster data, or that the open cluster mass functions are already influenced by dynamical effects. 
Detailed $N$-body modeling of the depicted clusters and their dynamical evolution in the Milky Way might therefore be necessary
to test if the open clusters formed with their present-day mass functions.

\section{Conclusions} \label{sec:conclusions}

We have determined the stellar mass functions of 91 Milky Way globular clusters and 29 massive star clusters in the Large and Small Magellanic Clouds
by fitting $N$-body models to star count data derived from over 300 individual HST fields. We find that the stellar mass functions of dynamically unevolved 
star clusters, characterized by relaxation times $T_{RH}$ of the order of their ages and/or lifetimes $T_{Diss}$ significantly larger than their ages,
are well described by a multi-stage power-law mass function $N(m) \sim m^{\alpha}$ with break masses at 0.4 M$_\odot$ and 1 M$_\odot$
and power law-slopes of $\alpha_{Low} =-0.3 \pm 0.2$, $\alpha_{Med}=-1.65 \pm 0.20$ and $\alpha_{High}=-2.30 \pm 0.20$ for the low-mass, intermediate-mass and high-mass stars respectively. 
An alternative description of the mass functions of these clusters is a log-normal mass function with characteristic mass $\log M_C = -0.36$ and width $\sigma_C=0.28$ that 
transitions into a power-law mass function at 1 M$_\odot$.
The mass function we find has fewer low-mass stars compared to the mass functions suggested by \cite{kroupa2001} or \citet{chabrier2003}, which are measured in the solar neighborhood 
and are significantly more bottom heavy. Our results therefore add to the increasing evidence that the stellar mass function varies across different environments.

We also find that star clusters with relaxation times or lifetimes much less than their ages are depleted in low-mass stars. The amount of depletion is
in agreement with $N$-body simulations that model the effects of mass segregation and preferential depletion of low-mass stars due to the external tidal fields on
star clusters.
Most investigated star clusters are therefore compatible with having formed with the same stellar mass function. By comparing the location of clusters in the dynamical age   
and age over lifetime vs. mass function slope planes with the results of direct $N$-body simulations, we also find that the mass function changes are best reproduced if most
black holes that form in star clusters receive strong natal kicks at birth that remove them from their parent clusters. The reason for this is that clusters which 
retain most of their black holes do not become sufficiently depleted in low-mass stars to explain the observations. We predict a remaining black hole population in star 
clusters of no more than a few hundred black holes for clusters with masses up to $10^6$ M$_\odot$. This estimate agrees with what has been found earlier from an analysis of the surface 
density profiles of globular clusters \citep{askaretal2018} and their amount of mass segregation \citep{weatherfordetal2018}.

Our data finally argues against IMF variations with either the metallicity, mass, density or velocity dispersion of star clusters, at least among relatively old stars and for metallicities
below the solar one. \citep{chonetal2021} argued, based on hydrodynamical simulations of star formation, for a shift towards more bottom-heavy mass functions
in higher metallicity environments. If real this shift must happen at metallicities $[Fe/H]<-2.3$ which is not probed by our data.

A mass function described by our findings leads to an average stellar mass at birth of $\langle m \rangle=1.0$ M$_\odot$ and an about twice larger fraction of 
massive stars with $m_*>10$~M$_\odot$ that evolve into black holes and neutron stars per unit stellar mass compared to a \citet{kroupa2001} mass function. 
Furthermore, at birth such a mass function has a nearly twice as large fraction of mass in $m>10$~M$_\odot$ stars compared to $m<0.8$ M$_\odot$ stars than
a \citet{kroupa2001} mass function.
Our results therefore argue for more self-enrichment of stars in globular clusters and, if they can be generalized to
field stars, overall more chemical enrichment in low metallicity environments. Due to the larger mass fraction in high-mass stars which can pollute
lower mass stars forming in the same cluster, our mass function would also help alleviate
the so-called mass budget problem in globular clusters \citep{dantonacaloi2008}, according to which the observed number of chemically enriched, second population stars is far higher 
than expected based on standard pollution scenarios. A factor two increase in the mass ratio of massive to low-mass stars is however not enough to explain
the observed number of second population stars since proposed scenarios for the origin of second generation stars fail by at least a factor of 10 to explain the 
current number ratios \citep{decressinetal2007,conroy2012}. Furthermore, these calculations assume that all of the the ejecta of massive stars are being used to enrich 
low-mass stars, which seems an optimistic assumption given that some of this material might escape or enrich intermediate-mass stars.

A mass function described by our findings also increases the amount of gravitational waves
created by inspiraling black holes and neutron stars compared to a standard mass function \citep{weatherfordetal2021}. 
However the number of inspiraling black holes also depends strongly on the binary and higher order multiple properties of massive stars 
\citep[e.g.][]{belczynskietal2016,antoninietal2017} and we currently do not have good constraints on these.
An independent test of our results can be obtained from stellar kinematics since the large number of compact remnants predicted by the mass function
found here should lead to an increase of stellar velocities over that predicted by standard mass functions with fewer remnants. We will investigate 
this point for a number of well observed globular clusters in a companion paper \citep{dicksonetal2023}.

\section*{Acknowledgments}

We dedicate this paper to the memory of our friend and long-term collaborator Antonio Sollima who passed away prior to the publication of this paper.
Antonio was a kind and humble scientist who made several key contributions to star cluster research, and he will be greatly missed.
We thank Emanuele Dalessandro, Hamid Ebrahimi, Mojyaba Taheri, Andr\'es del Pino, Sven Martens and Giacomo Cordoni for sharing their kinematic and mass function data with us. 
We also thank an anonymous referee for comments that improved the presentation of the paper.
VHB acknowledges the support of the Natural Sciences and Engineering Research Council of Canada (NSERC) through grant RGPIN-2020-05990. ND is grateful for the support of the Durland Scholarship in Graduate Research.
This work is based on observations made with the NASA/ESA Hubble Space Telescope, obtained from the data archive at the Space Telescope Science Institute. STScI is operated by the Association of Universities for Research in Astronomy, Inc. under NASA contract NAS 5-26555. Part of this work was performed on the OzSTAR 
national facility at Swinburne University of Technology. The OzSTAR program receives funding in part from the Astronomy National Collaborative Research 
Infrastructure Strategy (NCRIS) allocation provided by the Australian Government.\\[-0.4cm]

\section*{Data Availability}

Data is available upon request.\\[-0.4cm]

\bibliographystyle{mn2e}
\bibliography{mybib}

\label{lastpage}

\vspace*{-0.5cm}
\appendix
\section{Finding charts of the HST fields}

The HST finding charts are available as supplementary material  in the electronic version.

\section{Adopted parameters and derived mass function slopes for the clusters studied in this paper}
\begin{table*}
\caption{The adopted ages and metallicities, the best-fitting distance moduli and reddening values, the lower and upper mass limits of the mass function fits, and the derived mass function slopes for the globular clusters studied in this paper.
 The final three columns give $\alpha_{Tot}$, the best-fitting single power-law slope fit over the whole mass range, as well as the mass function slopes $\alpha_{Low}$ and $\alpha_{Med}$ for stars with masses below and above 0.40 M$_\odot$.}
\footnotesize
\begin{tabular}{lccccccccc}
\hline
\multirow{2}{*}{Cluster} & Age &  \multirow{2}{*}{[Fe/H]} & Dist. & \multirow{2}{*}{E(B-V)} & $m_{Low}$ & $m_{High}$  & \multirow{2}{*}{$\alpha_{Tot}$} & \multirow{2}{*}{$\alpha_{Low}$} & \multirow{2}{*}{$\alpha_{Med}$}\\ 
        &  [Gyr] & & mod. & & [M$_\odot$] &  [M$_\odot$] &  &  & \\ 
\hline
  \\[-0.2cm] 
Arp 2      &    12.4 & -1.74 & 17.29 &  0.14 &  0.41 &  0.80 &  $  -0.75 \pm 0.14$ &  ---  &  $  -0.75 \pm 0.14$ \\[+0.03cm] 
E 3        &    12.1 & -0.73 & 14.18 &  0.37 &  0.35 &  0.85 &  $ +0.88 \pm 0.24$ &  ---  &  $ +0.40 \pm 0.31$ \\[+0.03cm] 
IC 4499    &    12.1 & -1.62 & 16.38 &  0.26 &  0.28 &  0.81 &  $  -1.23 \pm 0.15$ & $ -0.07 \pm 0.32 $ &  $  -1.80 \pm 0.16$ \\[+0.03cm] 
Lynga 7    &    13.2 & -1.01 & 14.39 &  0.80 &  0.32 &  0.80 &  $ +0.30 \pm 0.16$ &  ---  &  $ +0.39 \pm 0.27$ \\[+0.03cm] 
NGC 104    &    12.4 & -0.76 & 13.29 &  0.05 &  0.22 &  0.86 &  $  -0.65 \pm 0.12$ & $ -0.25 \pm 0.17 $ &  $  -1.33 \pm 0.26$ \\[+0.03cm] 
NGC 288    &    11.7 & -1.32 & 14.83 &  0.04 &  0.22 &  0.82 &  $  -0.66 \pm 0.12$ & $ -0.51 \pm 0.12 $ &  $  -0.98 \pm 0.35$ \\[+0.03cm] 
NGC 362    &    11.0 & -1.30 & 14.80 &  0.03 &  0.26 &  0.82 &  $  -0.76 \pm 0.13$ &  ---  &  $  -0.91 \pm 0.25$ \\[+0.03cm] 
NGC 1261   &    11.0 & -1.27 & 16.07 &  0.02 &  0.33 &  0.83 &  $  -0.61 \pm 0.13$ & $ -0.36 \pm 0.16 $ &  $  -1.02 \pm 0.17$ \\[+0.03cm] 
NGC 1851   &    11.3 & -1.18 & 15.39 &  0.04 &  0.20 &  0.83 &  $  -0.70 \pm 0.13$ & $ -0.52 \pm 0.21 $ &  $  -0.91 \pm 0.15$ \\[+0.03cm] 
NGC 2298   &    13.1 & -1.96 & 14.96 &  0.23 &  0.25 &  0.77 &  $ +0.17 \pm 0.13$ & $ +0.31 \pm 0.16 $ &  $  -0.07 \pm 0.22$ \\[+0.03cm] 
NGC 2808   &    11.2 & -1.14 & 15.07 &  0.21 &  0.22 &  0.83 &  $  -0.60 \pm 0.14$ & $ -0.27 \pm 0.26 $ &  $  -1.19 \pm 0.20$ \\[+0.03cm] 
NGC 3201   &    11.2 & -1.51 & 13.38 &  0.29 &  0.24 &  0.80 &  $  -1.02 \pm 0.16$ & $ -0.86 \pm 0.19 $ &  $  -1.51 \pm 0.34$ \\[+0.03cm] 
NGC 4147   &    12.5 & -1.78 & 16.39 &  0.02 &  0.20 &  0.77 &  $  -0.05 \pm 0.13$ & $ +0.28 \pm 0.15 $ &  $  -0.09 \pm 0.22$ \\[+0.03cm] 
NGC 4372   &    13.0 & -2.19 & 13.43 &  0.53 &  0.16 &  0.77 &  $  -0.86 \pm 0.18$ & $ -0.59 \pm 0.40 $ &  $  -1.20 \pm 0.38$ \\[+0.03cm] 
NGC 4590   &    12.2 & -2.27 & 15.09 &  0.07 &  0.18 &  0.77 &  $  -1.18 \pm 0.14$ & $ -1.01 \pm 0.23 $ &  $  -1.72 \pm 0.17$ \\[+0.03cm] 
NGC 4833   &    12.9 & -1.89 & 14.11 &  0.36 &  0.24 &  0.77 &  $  -0.34 \pm 0.15$ & $ -0.04 \pm 0.38 $ &  $  -0.73 \pm 0.22$ \\[+0.03cm] 
NGC 5024   &    12.7 & -2.06 & 16.37 &  0.03 &  0.26 &  0.78 &  $  -1.20 \pm 0.12$ & $ -0.59 \pm 0.25 $ &  $  -1.63 \pm 0.21$ \\[+0.03cm] 
NGC 5053   &    12.8 & -2.30 & 16.27 &  0.01 &  0.23 &  0.78 &  $  -1.09 \pm 0.14$ & $ -0.73 \pm 0.35 $ &  $  -2.03 \pm 0.38$ \\[+0.03cm] 
NGC 5139   &    13.2 & -1.64 & 13.67 &  0.13 &  0.17 &  0.80 &  $  -0.80 \pm 0.11$ & $ -0.28 \pm 0.15 $ &  $  -1.61 \pm 0.28$ \\[+0.03cm] 
NGC 5272   &    12.1 & -1.50 & 15.04 &  0.01 &  0.17 &  0.77 &  $  -1.01 \pm 0.12$ & $ -0.78 \pm 0.37 $ &  $  -1.45 \pm 0.19$ \\[+0.03cm] 
NGC 5286   &    12.8 & -1.70 & 15.23 &  0.24 &  0.37 &  0.75 &  $  -0.73 \pm 0.16$ &  ---  &  $  -0.93 \pm 0.20$ \\[+0.03cm] 
NGC 5466   &    12.6 & -2.31 & 16.07 &  0.03 &  0.21 &  0.78 &  $  -0.86 \pm 0.12$ & $ -0.76 \pm 0.14 $ &  $  -0.97 \pm 0.25$ \\[+0.03cm] 
NGC 5897   &    12.3 & -1.90 & 15.49 &  0.14 &  0.24 &  0.79 &  $  -1.01 \pm 0.11$ & $ -0.68 \pm 0.21 $ &  $  -1.43 \pm 0.14$ \\[+0.03cm] 
NGC 5904   &    11.7 & -1.33 & 14.37 &  0.05 &  0.16 &  0.79 &  $  -0.76 \pm 0.11$ & $ -0.50 \pm 0.15 $ &  $  -1.22 \pm 0.14$ \\[+0.03cm] 
NGC 5927   &    11.6 & -0.29 & 14.49 &  0.43 &  0.50 &  0.81 &  $  -0.99 \pm 0.15$ &  ---  &  $  -0.99 \pm 0.15$ \\[+0.03cm] 
NGC 5946   &    12.0 & -1.29 & 15.13 &  0.60 &  0.52 &  0.80 &  $  -0.64 \pm 0.16$ &  ---  &  $  -0.64 \pm 0.16$ \\[+0.03cm] 
NGC 5986   &    12.5 & -1.63 & 15.11 &  0.31 &  0.43 &  0.76 &  $  -0.59 \pm 0.17$ & $ -0.46 \pm 0.36 $ &  $  -1.06 \pm 0.20$ \\[+0.03cm] 
NGC 6093   &    13.0 & -1.75 & 15.13 &  0.21 &  0.23 &  0.76 &  $  -0.45 \pm 0.16$ & $ -0.04 \pm 0.19 $ &  $  -0.46 \pm 0.24$ \\[+0.03cm] 
NGC 6101   &    12.6 & -1.98 & 15.83 &  0.13 &  0.24 &  0.79 &  $  -0.90 \pm 0.14$ & $ -0.38 \pm 0.23 $ &  $  -1.88 \pm 0.19$ \\[+0.03cm] 
NGC 6121   &    12.0 & -1.18 & 11.34 &  0.51 &  0.14 &  0.80 &  $  -0.34 \pm 0.15$ & $ -0.49 \pm 0.30 $ &  $  -0.73 \pm 0.33$ \\[+0.03cm] 
NGC 6144   &    13.2 & -1.82 & 14.51 &  0.41 &  0.36 &  0.76 &  $ +0.07 \pm 0.14$ & $ +0.24 \pm 0.17 $ &  $ +0.07 \pm 0.25$ \\[+0.03cm] 
NGC 6171   &    13.4 & -1.03 & 13.70 &  0.43 &  0.23 &  0.78 &  $  -0.29 \pm 0.14$ & $ -0.54 \pm 0.32 $ &  $  -0.11 \pm 0.30$ \\[+0.03cm] 
NGC 6205   &    12.6 & -1.58 & 14.45 &  0.01 &  0.30 &  0.74 &  $  -0.94 \pm 0.12$ &  ---  &  $  -1.36 \pm 0.15$ \\[+0.03cm] 
NGC 6218   &    13.3 & -1.33 & 13.54 &  0.21 &  0.20 &  0.78 &  $  -0.31 \pm 0.15$ & $ -0.38 \pm 0.17 $ &  $  -0.18 \pm 0.28$ \\[+0.03cm] 
NGC 6254   &    12.1 & -1.57 & 13.63 &  0.29 &  0.22 &  0.78 &  $  -0.60 \pm 0.13$ & $ -0.49 \pm 0.36 $ &  $  -1.02 \pm 0.18$ \\[+0.03cm] 
NGC 6266   &    12.3 & -1.18 & 14.03 &  0.49 &  0.53 &  0.81 &  $  -1.14 \pm 0.19$ &  ---  &  $  -1.14 \pm 0.19$ \\[+0.03cm] 
NGC 6273   &    12.4 & -1.76 & 14.61 &  0.44 &  0.43 &  0.76 &  $  -1.33 \pm 0.20$ &  ---  &  $  -1.33 \pm 0.20$ \\[+0.03cm] 
NGC 6284   &    11.8 & -1.31 & 15.76 &  0.33 &  0.43 &  0.82 &  $  -0.05 \pm 0.16$ &  ---  &  $  -0.05 \pm 0.16$ \\[+0.03cm] 
NGC 6287   &    13.9 & -2.12 & 14.50 &  0.72 &  0.36 &  0.76 &  $  -0.57 \pm 0.16$ &  ---  &  $  -0.96 \pm 0.38$ \\[+0.03cm] 
NGC 6293   &    12.8 & -2.01 & 14.82 &  0.40 &  0.24 &  0.78 &  $  -0.50 \pm 0.20$ & $ +0.41 \pm 0.48 $ &  $  -0.80 \pm 0.29$ \\[+0.03cm] 
NGC 6304   &    12.0 & -0.37 & 13.94 &  0.52 &  0.43 &  0.89 &  $  -0.40 \pm 0.14$ &  ---  &  $  -0.40 \pm 0.14$ \\[+0.03cm] 
NGC 6333   &    12.6 & -1.79 & 14.60 &  0.38 &  0.36 &  0.78 &  $  -0.39 \pm 0.21$ &  ---  &  $  -0.46 \pm 0.36$ \\[+0.03cm] 
NGC 6341   &    13.0 & -2.35 & 14.65 &  0.02 &  0.15 &  0.77 &  $  -0.83 \pm 0.12$ & $ -0.41 \pm 0.12 $ &  $  -1.35 \pm 0.20$ \\[+0.03cm] 
NGC 6342   &    12.5 & -0.49 & 14.52 &  0.53 &  0.41 &  0.82 &  $ +0.65 \pm 0.17$ &  ---  &  $ +0.65 \pm 0.17$ \\[+0.03cm] 
NGC 6352   &    11.8 & -0.62 & 13.67 &  0.30 &  0.23 &  0.83 &  $  -0.37 \pm 0.16$ & $ -0.35 \pm 0.50 $ &  $  -0.16 \pm 0.24$ \\[+0.03cm] 
NGC 6355   &    13.2 & -1.33 & 14.69 &  0.87 &  0.50 &  0.78 &  $  -0.50 \pm 0.19$ &  ---  &  $  -0.50 \pm 0.19$ \\[+0.03cm] 
NGC 6362   &    12.7 & -1.07 & 14.42 &  0.08 &  0.21 &  0.80 &  $  -0.49 \pm 0.15$ & $ -0.59 \pm 0.35 $ &  $  -0.69 \pm 0.23$ \\[+0.03cm] 
NGC 6366   &    11.7 & -0.59 & 12.68 &  0.76 &  0.39 &  0.83 &  $  -0.32 \pm 0.17$ & $ -0.41 \pm 0.44 $ &  $  -0.45 \pm 0.32$ \\[+0.03cm] 
NGC 6388   &    10.9 & -0.45 & 15.34 &  0.34 &  0.40 &  0.84 &  $  -1.01 \pm 0.18$ &  ---  &  $  -1.01 \pm 0.18$ \\[+0.03cm] 
NGC 6397   &    13.3 & -1.99 & 11.97 &  0.24 &  0.34 &  0.78 &  $  -0.32 \pm 0.14$ & $ -0.43 \pm 0.27 $ &  $  -0.92 \pm 0.32$ \\[+0.03cm] 
NGC 6401   &    13.2 & -1.01 & 14.13 &  0.96 &  0.40 &  0.81 &  $  -0.37 \pm 0.17$ &  ---  &  $  -0.37 \pm 0.17$ \\[+0.03cm] 
NGC 6402   &    12.0 & -1.28 & 14.80 &  0.60 &  0.38 &  0.81 &  $  -0.81 \pm 0.15$ &  ---  &  $  -0.84 \pm 0.20$ \\[+0.03cm] 
NGC 6426   &    13.1 & -2.36 & 16.58 &  0.43 &  0.24 &  0.78 &  $  -1.07 \pm 0.14$ & $ -0.41 \pm 0.18 $ &  $  -1.59 \pm 0.15$ \\[+0.03cm] 
NGC 6496   &    11.4 & -0.44 & 14.91 &  0.16 &  0.31 &  0.82 &  $  -0.25 \pm 0.16$ & $ +0.11 \pm 0.44 $ &  $  -0.18 \pm 0.30$ \\[+0.03cm] 
NGC 6535   &    12.6 & -1.79 & 14.12 &  0.48 &  0.29 &  0.79 &  $ +1.13 \pm 0.18$ &  ---  &  $ +0.92 \pm 0.38$ \\[+0.03cm]
\hline
\end{tabular}
\label{table:main}
\end{table*}

\begin{table*}
\contcaption{}
\footnotesize
\begin{tabular}{lccccccccc}
\hline
\multirow{2}{*}{Cluster} & Age &  \multirow{2}{*}{[Fe/H]} & Dist. & \multirow{2}{*}{E(B-V)} & $m_{Low}$ & $m_{High}$  & \multirow{2}{*}{$\alpha_{Tot}$} & \multirow{2}{*}{$\alpha_{Low}$} & \multirow{2}{*}{$\alpha_{Med}$}\\ 
        &  [Gyr] & & mod. & & [M$_\odot$] &  [M$_\odot$] &  &  & \\ 
\hline
  \\[-0.2cm] 
NGC 6541   &    12.9 & -1.82 & 14.44 &  0.16 &  0.26 &  0.73 &  $  -0.75 \pm 0.16$ & $ -0.60 \pm 0.51 $ &  $  -1.17 \pm 0.22$ \\[+0.03cm]
NGC 6544   &    11.1 & -1.47 & 12.16 &  0.82 &  0.38 &  0.82 &  $  -0.49 \pm 0.19$ &  ---  &  $  -0.46 \pm 0.26$ \\[+0.03cm]
NGC 6558   &    12.3 & -1.37 & 14.67 &  0.40 &  0.37 &  0.80 &  $ +1.53 \pm 0.15$ &  ---  &  $ +1.49 \pm 0.16$ \\[+0.03cm]
NGC 6584   &    12.0 & -1.50 & 15.62 &  0.11 &  0.30 &  0.78 &  $  -0.80 \pm 0.13$ & $ -0.61 \pm 0.13 $ &  $  -0.95 \pm 0.19$ \\[+0.03cm]
NGC 6624   &    11.9 & -0.42 & 14.52 &  0.27 &  0.28 &  0.82 &  $ +0.66 \pm 0.16$ &  ---  &  $ +0.48 \pm 0.17$ \\[+0.03cm] 
NGC 6626   &    12.7 & -1.46 & 13.64 &  0.49 &  0.19 &  0.78 &  $  -0.18 \pm 0.18$ & $ -0.24 \pm 0.40 $ &  $  -0.31 \pm 0.19$ \\[+0.03cm] 
NGC 6637   &    11.8 & -0.59 & 14.75 &  0.20 &  0.26 &  0.86 &  $ +0.10 \pm 0.19$ &  ---  &  $  -0.13 \pm 0.29$ \\[+0.03cm] 
NGC 6638   &    12.0 & -0.99 & 15.05 &  0.40 &  0.42 &  0.82 &  $ +0.71 \pm 0.22$ &  ---  &  $ +0.71 \pm 0.22$ \\[+0.03cm] 
NGC 6642   &    12.7 & -1.19 & 14.43 &  0.44 &  0.32 &  0.80 &  $ +1.00 \pm 0.15$ &  ---  &  $ +0.87 \pm 0.18$ \\[+0.03cm] 
NGC 6652   &    11.9 & -0.76 & 14.88 &  0.15 &  0.26 &  0.83 &  $ +0.84 \pm 0.17$ & $ +1.11 \pm 0.51 $ &  $ +0.90 \pm 0.33$ \\[+0.03cm] 
NGC 6656   &    12.8 & -1.70 & 12.66 &  0.39 &  0.18 &  0.74 &  $  -0.90 \pm 0.15$ & $ -0.58 \pm 0.35 $ &  $  -1.22 \pm 0.25$ \\[+0.03cm] 
NGC 6681   &    13.0 & -1.62 & 14.86 &  0.16 &  0.23 &  0.78 &  $  -0.17 \pm 0.13$ & $ -0.41 \pm 0.45 $ &  $ +0.20 \pm 0.15$ \\[+0.03cm] 
NGC 6712   &    11.2 & -1.02 & 14.24 &  0.55 &  0.35 &  0.84 &  $  -0.24 \pm 0.17$ &  ---  &  $  -0.10 \pm 0.28$ \\[+0.03cm] 
NGC 6715   &    12.0 & -1.44 & 17.15 &  0.16 &  0.52 &  0.80 &  $  -1.43 \pm 0.19$ &  ---  &  $  -1.43 \pm 0.19$ \\[+0.03cm] 
NGC 6717   &    12.8 & -1.26 & 14.38 &  0.23 &  0.30 &  0.79 &  $ +1.47 \pm 0.18$ &  ---  &  $ +1.29 \pm 0.36$ \\[+0.03cm] 
NGC 6723   &    12.8 & -1.10 & 14.54 &  0.11 &  0.23 &  0.79 &  $  -0.33 \pm 0.14$ & $ -0.14 \pm 0.24 $ &  $  -0.58 \pm 0.21$ \\[+0.03cm] 
NGC 6752   &    12.8 & -1.55 & 13.09 &  0.09 &  0.14 &  0.78 &  $  -0.67 \pm 0.13$ & $ -0.61 \pm 0.27 $ &  $  -0.88 \pm 0.32$ \\[+0.03cm] 
NGC 6779   &    13.2 & -2.00 & 15.19 &  0.25 &  0.31 &  0.77 &  $  -0.56 \pm 0.15$ & $ -0.41 \pm 0.20 $ &  $  -0.84 \pm 0.18$ \\[+0.03cm] 
NGC 6809   &    13.3 & -1.93 & 13.61 &  0.15 &  0.16 &  0.67 &  $  -0.93 \pm 0.13$ & $ -0.81 \pm 0.16 $ &  $  -0.86 \pm 0.24$ \\[+0.03cm] 
NGC 6838   &    11.9 & -0.82 & 13.01 &  0.28 &  0.25 &  0.82 &  $  -0.17 \pm 0.15$ & $ -0.06 \pm 0.20 $ &  $  -0.80 \pm 0.30$ \\[+0.03cm] 
NGC 6934   &    12.0 & -1.56 & 15.93 &  0.12 &  0.38 &  0.79 &  $  -1.15 \pm 0.15$ &  ---  &  $  -1.26 \pm 0.15$ \\[+0.03cm] 
NGC 6981   &    11.9 & -1.48 & 16.11 &  0.06 &  0.40 &  0.81 &  $  -0.68 \pm 0.13$ &  ---  &  $  -0.68 \pm 0.13$ \\[+0.03cm] 
NGC 7006   &    12.2 & -1.46 & 17.97 &  0.09 &  0.36 &  0.79 &  $  -1.26 \pm 0.14$ &  ---  &  $  -1.52 \pm 0.16$ \\[+0.03cm] 
NGC 7078   &    13.0 & -2.33 & 15.15 &  0.11 &  0.18 &  0.73 &  $  -0.97 \pm 0.12$ & $ -0.57 \pm 0.28 $ &  $  -1.36 \pm 0.15$ \\[+0.03cm] 
NGC 7089   &    12.1 & -1.66 & 15.44 &  0.06 &  0.24 &  0.78 &  $  -0.87 \pm 0.16$ & $ -0.62 \pm 0.35 $ &  $  -1.17 \pm 0.19$ \\[+0.03cm] 
NGC 7099   &    13.1 & -2.33 & 14.67 &  0.07 &  0.20 &  0.77 &  $  -0.76 \pm 0.13$ & $ -0.73 \pm 0.16 $ &  $  -1.12 \pm 0.16$ \\[+0.03cm] 
Pal 1      &  \,    7.6 & -0.51 & 15.24 &  0.20 &  0.20 &  1.00 &  $  -0.51 \pm 0.17$ & $ -0.04 \pm 0.34 $ &  $  -1.15 \pm 0.28$ \\[+0.03cm] 
Pal 5      &    11.1 & -1.41 & 16.66 &  0.11 &  0.29 &  0.81 &  $  -0.84 \pm 0.19$ &   $ -0.39 \pm 0.44 $  &  $  -0.96 \pm 0.27$ \\[+0.03cm] 
Pal 12     &  \,    9.3 & -0.81 & 16.33 &  0.02 &  0.37 &  0.84 &  $  -0.61 \pm 0.15$ & --- &  $  -0.76 \pm 0.19$ \\[+0.03cm] 
Pal 13     &    13.4 & -1.78 & 16.70 &  0.18 &  0.24 &  0.78 &  $ +0.29 \pm 0.22$ & $ +0.93 \pm 0.31 $ &  $  -0.12 \pm 0.40$ \\[+0.03cm] 
Pal 15     &    12.7 & -2.10 & 18.22 &  0.45 &  0.47 &  0.78 &  $  -1.41 \pm 0.16$ &  ---  &  $  -1.41 \pm 0.16$ \\[+0.03cm] 
Pyxis      &    11.2 & -1.20 & 17.81 &  0.28 &  0.42 &  0.83 &  $  -1.44 \pm 0.16$ &  ---  &  $  -1.44 \pm 0.16$ \\[+0.03cm] 
Rup 106    &    11.2 & -1.78 & 16.58 &  0.20 &  0.22 &  0.81 &  $  -0.81 \pm 0.14$ & $ -0.35 \pm 0.19 $ &  $  -1.11 \pm 0.15$ \\[+0.03cm] 
Sgr II     &    12.0 & -2.28 & 18.92 &  0.22 &  0.45 &  0.80 &  $  -1.51 \pm 0.17$ &  ---  &  $  -1.51 \pm 0.17$ \\[+0.03cm] 
Ter 7      &  \,    7.3 & -0.12 & 16.93 &  0.05 &  0.35 &  1.04 &  $  -1.61 \pm 0.14$ &  ---  &  $  -1.85 \pm 0.16$ \\[+0.03cm] 
Ter 8      &    13.2 & -2.16 & 17.20 &  0.16 &  0.28 &  0.77 &  $  -1.14 \pm 0.15$ &  ---  &  $  -1.92 \pm 0.20$ \\[+0.03cm] 
\hline
\end{tabular}
\end{table*}

\begin{table*}
\caption{Adopted parameters and derived mass function slopes for the LMC/SMC star clusters. The final column gives the mass function slope of the high-mass stars with $m>1$ M$_\odot$. The meaning of the other columns is the same as in Table~\ref{table:main}.}
\begin{tabular}{lccccccccc}
\hline
\multirow{2}{*}{Cluster} & log Age & \multirow{2}{*}{[Fe/H]} & Dist. & \multirow{2}{*}{E(B-V)} & $m_{Low}$ & $m_{High}$ & \multirow{2}{*}{$\alpha_{Tot}$} & \multirow{2}{*}{$\alpha_{Med}$} & \multirow{2}{*}{$\alpha_{High}$}\\ 
        &  [yr] & & mod. & & [M$_\odot$] &  [M$_\odot$] & & & \\ 
\hline
  \\[-0.2cm] 
Fornax 1   &   10.08 &  0.00 & 20.73 &  0.05 &  0.67 &  0.79 & $-1.19 \pm 0.14$ & $ -1.18 \pm 0.14 $ &  ---  \\[+0.03cm] 
Fornax 3   &   10.08 & -2.33 & 20.73 &  0.03 &  0.66 &  0.79 & $-1.15 \pm 0.12$ & $ -1.15 \pm 0.20 $ &  ---  \\[+0.03cm] 
Hodge 6    & $\,\,$  9.40 & -0.35 & 18.50 &  0.09 &  0.75 &  1.37 & $-2.25 \pm 0.13$ & $ -1.39 \pm 0.22 $ & $ -2.48 \pm 0.21 $ \\[+0.03cm] 
Hodge 301  & $\,\,$  7.38 & -0.30 & 18.50 &  0.10 &  1.28 &  6.26 & $-2.36 \pm 0.24$ &  ---  & $ -2.36 \pm 0.24 $ \\[+0.03cm] 
Kron 3     & $\,\,$  9.81 & -1.08 & 18.91 &  0.03 &  0.49 &  0.98 & $-1.15 \pm 0.12$ & $ -1.15 \pm 0.12 $ &  ---  \\[+0.03cm] 
Lindsay  1 & $\,\,$  9.88 & -1.14 & 18.78 &  0.06 &  0.48 &  0.94 & $-1.19 \pm 0.14$ & $ -1.18 \pm 0.14 $ &  ---  \\[+0.03cm] 
Lindsay  38 & $\,\,$  9.81 & -1.59 & 19.12 &  0.02 &  0.49 &  0.94 & $-1.46 \pm 0.22$ & $ -1.47 \pm 0.22 $ &  ---  \\[+0.03cm] 
Lindsay  113 & $\,\,$  9.72 & -1.44 & 18.80 &  0.00 &  0.54 &  0.99 & $-0.80 \pm 0.19$ & $ -1.24 \pm 0.21 $ &  ---  \\[+0.03cm] 
NGC 121    &   10.02 & -1.46 & 19.06 &  0.05 &  0.53 &  0.84 & $-0.96 \pm 0.18$ & $ -1.16 \pm 0.18 $ &  ---  \\[+0.03cm] 
NGC 330    & $\,\,$  7.49 & -0.98 & 18.90 &  0.07 &  0.78 &  6.96 & $-2.38 \pm 0.25$ &  ---  & $ -2.18 \pm 0.25 $ \\[+0.03cm] 
NGC 339    & $\,\,$  9.78 & -1.14 & 18.80 &  0.03 &  0.54 &  1.01 & $-1.66 \pm 0.13$ & $ -1.66 \pm 0.13 $ &  ---  \\[+0.03cm] 
NGC 416    & $\,\,$  9.78 & -1.00 & 18.91 &  0.08 &  0.59 &  0.99 & $-1.46 \pm 0.21$ & $ -1.55 \pm 0.21 $ &  ---  \\[+0.03cm] 
NGC 419    & $\,\,$  9.18 & -0.67 & 18.85 &  0.02 &  0.53 &  1.50 & $-2.03 \pm 0.19$ & $ -2.03 \pm 0.22 $ &  ---  \\[+0.03cm] 
NGC 1651   & $\,\,$  9.30 & -0.53 & 18.46 &  0.02 &  0.58 &  1.37 & $-1.87 \pm 0.21$ & $ -1.21 \pm 0.15 $ & $ -2.05 \pm 0.15 $ \\[+0.03cm] 
NGC 1755   & $\,\,$  7.90 & -0.50 & 18.35 &  0.13 &  0.91 &  5.27 & $-2.37 \pm 0.27$ &  ---  & $ -2.58 \pm 0.29 $ \\[+0.03cm] 
NGC 1783   & $\,\,$  9.18 & -0.35 & 18.46 &  0.02 &  0.63 &  1.51 & $-1.93 \pm 0.19$ & $ -1.40 \pm 0.15 $ & $ -2.53 \pm 0.15 $ \\[+0.03cm] 
NGC 1806   & $\,\,$  9.18 & -0.60 & 18.46 &  0.02 &  0.60 &  1.57 & $-2.01 \pm 0.18$ &  ---  & $ -2.44 \pm 0.21 $ \\[+0.03cm] 
NGC 1846   & $\,\,$  9.18 & -0.49 & 18.46 &  0.06 &  0.65 &  1.66 & $-2.06 \pm 0.21$ &  ---  & $ -2.21 \pm 0.23 $ \\[+0.03cm] 
NGC 1850   & $\,\,$  8.00 & -0.31 & 18.35 &  0.10 &  0.67 &  4.56 & $-2.06 \pm 0.22$ &  ---  & $ -2.16 \pm 0.25 $ \\[+0.03cm] 
NGC 1856   & $\,\,$  8.54 & -0.30 & 18.44 &  0.15 &  0.88 &  2.96 & $-2.23 \pm 0.30$ &  ---  & $ -2.35 \pm 0.35 $ \\[+0.03cm] 
NGC 1866   & $\,\,$  8.23 & -0.36 & 18.50 &  0.09 &  0.73 &  3.93 & $-2.45 \pm 0.22$ &  ---  & $ -2.14 \pm 0.24 $ \\[+0.03cm] 
NGC 1978   & $\,\,$  9.30 & -0.35 & 18.40 &  0.05 &  0.62 &  1.44 & $-1.46 \pm 0.17$ & $ -1.10 \pm 0.17 $ & $ -2.04 \pm 0.17 $ \\[+0.03cm] 
NGC 2121   & $\,\,$  9.51 & -0.50 & 18.30 &  0.10 &  0.69 &  1.19 & $-2.19 \pm 0.29$ & $ -1.44 \pm 0.21 $ &  ---  \\[+0.03cm] 
NGC 2155   & $\,\,$  9.51 & -0.46 & 18.30 &  0.03 &  0.77 &  1.21 & $-0.60 \pm 0.53$ &  ---  &  ---  \\[+0.03cm] 
NGC 2173   & $\,\,$  9.20 & -0.42 & 18.44 &  0.10 &  0.63 &  1.58 & $-2.22 \pm 0.14$ &  ---  &  ---  \\[+0.03cm] 
NGC 2203   & $\,\,$  9.26 & -0.41 & 18.41 &  0.12 &  0.60 &  1.50 & $-1.41 \pm 0.16$ & $ -1.41 \pm 0.21 $ &  ---  \\[+0.03cm] 
R 136      & $\,\,$  6.20 & -0.30 & 18.50 &  0.40 &  1.82 & 35.03 & $-2.02 \pm 0.23$ &  ---  & $ -2.02 \pm 0.23 $ \\[+0.03cm] 
Reticulum  &   10.08 & -1.66 & 18.39 &  0.05 &  0.42 &  0.80 & $-1.28 \pm 0.16$ & $ -1.28 \pm 0.16 $ &  ---  \\[+0.03cm] 
SL 639     & $\,\,$  7.34 & -0.30 & 18.50 &  0.35 &  1.83 &  8.23 & $-2.48 \pm 0.26$ &  ---  & $ -2.48 \pm 0.26 $ \\[+0.03cm] 
\hline
\end{tabular}
\label{table:lmcsmc}
\end{table*}

\section{Sources of additional kinematic data of globular clusters}

\begin{table}
\caption{Sources of kinematic data on globular clusters used in this work in addition to \citet{baumgardt2017} and 
  \citet{baumgardthilker2018} (LOS = line-of-sight radial velocities, PM = proper motion velocity dispersions}
\begin{tabular}[h]{l@{\hspace{0.2cm}}l@{\hspace{0.2cm}}c}
\hline
Name & Source & Type \\ 
\hline
NGC 104  & \citet{cerniauskasetal2018} & LOS \\
         & \citet{libralatoetal2022}  & PM \\
         & \citet{martensetal2023} & LOS \\[+0.1cm]
NGC 288  & \citet{ferraroetal2018} & LOS \\
         & \citet{libralatoetal2022}  & PM \\[+0.1cm]
NGC 362  & \citet{ferraroetal2018} & LOS \\
         & \citet{libralatoetal2022} & PM \\
         & \citet{martensetal2023} & LOS \\[+0.1cm]
NGC 1261 & \citet{ferraroetal2018} & LOS  \\
         & \citet{munozetal2021} & LOS \\
         & \citet{marinoetal2021} & LOS \\
         & \citet{libralatoetal2022} & PM \\
         & \citet{wanetal2023} & LOS \\[+0.1cm]
NGC 1851 & \citet{ferraroetal2018} & LOS \\
         & \citet{libralatoetal2022} & PM \\
         & \citet{wanetal2023} & LOS \\
         & \citet{martensetal2023} & LOS \\[+0.1cm]
NGC 2298 & \citet{libralatoetal2022} & PM \\[+0.1cm]
NGC 2808 & \citet{martensetal2023} & LOS \\[+0.1cm]
NGC 3201 & \citet{ferraroetal2018} & LOS \\
         & \citet{giesersetal2019} & LOS \\
         & \citet{wanetal2021} & LOS \\
         & \citet{libralatoetal2022} & PM \\
         & \citet{martensetal2023} & LOS \\[+0.1cm]
NGC 4590 & \citet{libralatoetal2022} & PM \\[+0.1cm]
         & \citet{wanetal2023} & LOS \\[+0.1cm]
NGC 4833 & \citet{libralatoetal2022} & PM \\[+0.1cm]
NGC 5024 & \citet{libralatoetal2022} & PM \\[+0.1cm]
         & \citet{delpinoetal2022} & PM \\[+0.1cm]
NGC 5053 & \citet{delpinoetal2022} & PM \\[+0.1cm]
NGC 5139 & \citet{johnsonetal2020} & LOS \\[+0.1cm]
NGC 5272 & \citet{ferraroetal2018} & LOS \\
         & \citet{libralatoetal2022} & PM \\[+0.1cm]
NGC 5286 & \citet{libralatoetal2022} & PM \\
         & \citet{martensetal2023} & LOS \\[+0.1cm]
NGC 5466 & \citet{libralatoetal2022} & PM \\
         & \citet{delpinoetal2022} & PM \\[+0.1cm]
NGC 5897 & \citet{libralatoetal2022} & PM \\[+0.1cm]
NGC 5904 & \citet{libralatoetal2022} & PM \\
         & \citet{martensetal2023} & LOS \\[+0.1cm]
NGC 5927 & \citet{ferraroetal2018} & LOS \\
         & \citet{libralatoetal2022} & PM \\[+0.1cm]
NGC 6093 & \citet{goettgensetal2021} & LOS \\[+0.1cm]
NGC 6101 & \citet{libralatoetal2022} & PM \\[+0.1cm]
NGC 6121 & \citet{libralatoetal2022} & PM \\[+0.1cm]
NGC 6171 & \citet{ferraroetal2018} & LOS \\
         & \citet{libralatoetal2022} & PM \\[+0.1cm]
NGC 6205 & \citet{libralatoetal2022} & PM \\[+0.1cm]
NGC 6218 & \citet{libralatoetal2022} & PM \\
         & \citet{martensetal2023} & LOS \\[+0.1cm]
NGC 6254 & \citet{ferraroetal2018} & LOS \\
         & \citet{barthetal2020} & LOS \\
         & \citet{libralatoetal2022} & PM \\
         & \citet{martensetal2023} & LOS \\[+0.1cm]
\hline
\end{tabular}
\end{table}

\begin{table}
\contcaption{}
\begin{tabular}[h]{l@{\hspace{0.2cm}}l@{\hspace{0.2cm}}c}
\hline
Name & Source & Type \\
\hline
NGC 6266 & \citet{martensetal2023} & LOS \\[+0.1cm]
NGC 6293 & \citet{martensetal2023} & LOS \\[+0.1cm]
NGC 6304 & \citet{libralatoetal2022} & PM \\[+0.1cm]
NGC 6341 & \citet{libralatoetal2022} & PM \\[+0.1cm]
NGC 6342 & \citet{cohenetal2021} & PM \\[+0.1cm]
NGC 6352 & \citet{libralatoetal2022} & PM \\[+0.1cm]
NGC 6355 & \citet{cohenetal2021} & PM \\[+0.1cm]
NGC 6362 & \citet{dalessandroetal2021} & LOS \\
         & \citet{libralatoetal2022} & PM \\[+0.1cm]
NGC 6388 & \citet{libralatoetal2022} & PM \\
         & \citet{martensetal2023} & LOS \\[+0.1cm]
NGC 6397 & \citet{libralatoetal2022} & PM \\
         & \citet{martensetal2023} & LOS \\[+0.1cm]
NGC 6401 & \citet{cohenetal2021} & PM \\[+0.1cm]
NGC 6402 & \citet{johnsonetal2019} & LOS \\[+0.1cm]
NGC 6496 & \citet{ferraroetal2018} & LOS \\[+0.1cm]
NGC 6535 & \citet{libralatoetal2022} & PM \\[+0.1cm]
NGC 6541 & \citet{libralatoetal2022} & PM \\
         & \citet{martensetal2023} & LOS \\[+0.1cm]
NGC 6558 & \citet{cohenetal2021} & PM \\[+0.1cm]
NGC 6584 & \citet{libralatoetal2022} & PM \\[+0.1cm]
NGC 6624 & \citet{libralatoetal2022} & PM \\
         & \citet{martensetal2023} & LOS \\[+0.1cm]
NGC 6637 & \citet{libralatoetal2022} & PM \\[+0.1cm]
NGC 6642 & \citet{cohenetal2021} & PM \\[+0.1cm]
NGC 6652 & \citet{libralatoetal2022} & PM \\[+0.1cm]
NGC 6656 & \citet{libralatoetal2022} & PM \\
         & \citet{martensetal2023} & LOS \\[+0.1cm]
NGC 6681 & \citet{libralatoetal2022} & PM \\
         & \citet{martensetal2023} & LOS \\[+0.1cm]
NGC 6715 & \citet{libralatoetal2022} & PM \\
         & \citet{kacharovetal2022}  & LOS \\[+0.1cm]
NGC 6717 & \citet{libralatoetal2022} & PM \\[+0.1cm]
NGC 6723 & \citet{ferraroetal2018} & LOS  \\
         & \citet{crestanietal2019}  & LOS  \\
         & \citet{libralatoetal2022} & PM \\
         & \citet{taherietal2022} & PM \\[+0.1cm]
NGC 6752 & \citet{libralatoetal2022} & PM \\
         & \citet{martensetal2023} & LOS \\[+0.1cm]
NGC 6779 & \citet{libralatoetal2022} & PM \\[+0.1cm]
NGC 6809 & \citet{rainetal2019} & LOS \\
         & \citet{libralatoetal2022} & PM \\[+0.1cm]
NGC 6838 & \citet{barthetal2020} & LOS \\
         & \citet{libralatoetal2022} & PM \\[+0.1cm]
NGC 6934 & \citet{marinoetal2021} & LOS \\
         & \citet{libralatoetal2022} & PM \\[+0.1cm]
NGC 7078 & \citet{kirbyetal2020} & LOS \\
         & \citet{libralatoetal2022} & PM \\
         & \citet{martensetal2023} & LOS \\[+0.1cm]
NGC 7089 & \citet{libralatoetal2022} & PM \\
         & \citet{martensetal2023} & LOS \\[+0.1cm]
NGC 7099 & \citet{dalgleishetal2020} & LOS \\
         & \citet{libralatoetal2022} & PM \\
         & \citet{martensetal2023} & LOS \\[+0.1cm]
Pal 15   & \citet{kochetal2019} & LOS \\[+0.1cm]
\hline
\end{tabular}
\end{table}

\end{document}